\documentclass[twocolumn]{autart}    % Enable this line and disable the 
\usepackage{graphicx}          % Include this line if your 
\usepackage{mathtools}
\usepackage{amsmath}
\usepackage{units}

\usepackage{cite}
\usepackage{amssymb}
\usepackage{amsfonts}
\usepackage{mathrsfs}
\usepackage{csquotes}
\usepackage{subfig}
\usepackage{color}
\newcommand{\T}{\scriptscriptstyle\top}  

\usepackage{enumerate}
\usepackage{enumitem}
\DeclareUnicodeCharacter{2212}{\ensuremath{-}}

\newtheorem{problem}{Problem}
\newtheorem{corollary}{Corollary}

\makeatletter
\let\save@mathaccent\mathaccent
\newcommand*\if@single[3]{%
  \setbox0\hbox{${\mathaccent"0362{#1}}^H$}%
  \setbox2\hbox{${\mathaccent"0362{\kern0pt#1}}^H$}%
  \ifdim\ht0=\ht2 #3\else #2\fi
  }
\newcommand*\rel@kern[1]{\kern#1\dimexpr\macc@kerna}
\newcommand*\widebar[1]{\@ifnextchar^{{\wide@bar{#1}{0}}}{\wide@bar{#1}{1}}}
\newcommand*\wide@bar[2]{\if@single{#1}{\wide@bar@{#1}{#2}{1}}{\wide@bar@{#1}{#2}{2}}}
\newcommand*\wide@bar@[3]{%
  \begingroup
  \def\mathaccent##1##2{%
    \let\mathaccent\save@mathaccent
    \if#32 \let\macc@nucleus\first@char \fi
    \setbox\z@\hbox{$\macc@style{\macc@nucleus}_{}$}%
    \setbox\tw@\hbox{$\macc@style{\macc@nucleus}{}_{}$}%
    \dimen@\wd\tw@
    \advance\dimen@-\wd\z@
    \divide\dimen@ 3
    \@tempdima\wd\tw@
    \advance\@tempdima-\scriptspace
    \divide\@tempdima 10
    \advance\dimen@-\@tempdima
    \ifdim\dimen@>\z@ \dimen@0pt\fi
    \rel@kern{0.6}\kern-\dimen@
    \if#31
      \overline{\rel@kern{-0.6}\kern\dimen@\macc@nucleus\rel@kern{0.4}\kern\dimen@}%
      \advance\dimen@0.4\dimexpr\macc@kerna
      \let\final@kern#2%
      \ifdim\dimen@<\z@ \let\final@kern1\fi
      \if\final@kern1 \kern-\dimen@\fi
    \else
      \overline{\rel@kern{-0.6}\kern\dimen@#1}%
    \fi
  }%
  \macc@depth\@ne
  \let\math@bgroup\@empty \let\math@egroup\macc@set@skewchar
  \mathsurround\z@ \frozen@everymath{\mathgroup\macc@group\relax}%
  \macc@set@skewchar\relax
  \let\mathaccentV\macc@nested@a
  \if#31
    \macc@nested@a\relax111{#1}%
  \else
    \def\gobble@till@marker##1\endmarker{}%
    \futurelet\first@char\gobble@till@marker#1\endmarker
    \ifcat\noexpand\first@char A\else
      \def\first@char{}%
    \fi
    \macc@nested@a\relax111{\first@char}%
  \fi
  \endgroup
}

\usepackage{float}

\begin{document}
\allowdisplaybreaks

\begin{frontmatter}
%\runtitle{Insert a suggested running title}  % Running title for regular 
                                              % papers but only if the title  
                                              % is over 5 words. Running title 
                                              % is not shown in output.

\title{High-Performance Model Predictive Control
for Quadcopters with Formal Stability Guarantees}
%\thanksref{footnoteinfo}} % Title, preferably not more 
                                                % than 10 words.

%\thanks[footnoteinfo]{The research leading to these results has received funding from the European Research Council under the Advanced ERC Grant Agreement PROACTHIS, no. 101055384.}

%\author[Eindhoven]{Erfan Shakhesi}\ead{e.shakhesi@tue.nl},    % Add the 
%\author[Eindhoven]{W.P.M.H. (Maurice) Heemels}\ead{m.heemels@tue.nl},               % e-mail address 
%\author[Eindhoven]{Alexander Katriniok}\ead{a.katriniok@tue.nl}  % (ead) as shown

\author[Eindhoven]{Maedeh Izadi}
\ead{m.izadi.najafabadi1@tue.nl},
\author[Eindhoven]{A.T.J.R. Cobbenhagen}
\ead{a.t.j.r.cobbenhagen@tue.nl},
\author[Avular]{Ruben Sommer}
\ead{r.sommer@avular.com},
\author[Eindhoven]{A.R.P. Andrien}
\ead{a.andrien@gmail.com},
\author[Eindhoven]{Erjen Lefeber}
\ead{a.a.j.lefeber@tue.nl},
\author[Eindhoven]{W.P.M.H. (Maurice) Heemels}
\ead{m.heemels@tue.nl}

\address[Eindhoven]{Department of Mechanical Engineering, Eindhoven
University of Technology, The Netherlands.}  % 
\address[Avular]{Avular Mobile Robotics, Eindhoven, The Netherlands.}
          
\begin{keyword}                           % Five to ten keywords,  
Model predictive control, uniform global asymptotic stability, 
inner–outer loop control structure,
quadcopter.
\end{keyword}                             % keyword list or with the 
                                          % help of the Automatica 
                                          % keyword wizard

\begin{abstract}
In this paper, we present a novel cascade control structure with formal guarantees of uniform almost global asymptotic stability for the state tracking error dynamics of a quadcopter. The proposed approach features a model predictive control strategy for the outer loop, explicitly accounting for the non-zero total thrust constraint.
The outer-loop controller generates an acceleration reference, which is then converted into attitude, angular velocity and acceleration references, subsequently tracked by a nonlinear inner-loop controller. The proposed cascade control strategy is validated through numerical case studies, underlying high-fidelity models, demonstrating its ability to track fast trajectories with small error.
%Additionally, real-time experiments using the Avular Vertex One are conducted to validate the performance of the outer-loop MPC controller.
\end{abstract}

\end{frontmatter}
\section{Introduction} \label{sec:sec1}
\vspace{-0.3cm}
Quadcopters have become a prominent category of UAVs with a broad range of practical applications, including infrastructure inspection \cite{infrastructureapplication}, search-and-rescue missions \cite{searcrescueapplication}, agriculture \cite{aggricultureapplication},  
%aerial photography \cite{photographyapplication}, 
wildlife monitoring \cite{wildlifeapplication}, and medicine delivery \cite{medicineapplication}. This growing interest in quadcopters has driven the development of controllers capable of moving from laboratory demonstrations to real-world scenarios, calling for levels of high performance and robustness, while respecting the often limited on-board computational capabilities of quadcopters.

The control of quadcopters, especially trajectory tracking, is challenging due to the system's open-loop instability, nonlinear dynamics,
%3D geometric nonlinearity from rigid body dynamics,
and underactuated configuration. The differential flatness property of quadcopters \cite{flatnessbasedMPC} can be employed for trajectory planning to achieve a user-specified goal. Various strategies are used to address the trajectory tracking problem for quadrotors, including
sliding mode control \cite{slidingmode},
nonlinear PID control \cite{nonlinearpid},
iterative learning control \cite{iterativelearning},
nonlinear control \cite{ErijenNLcontroller}, reinforcement learning \cite{reinforcementlearning},
%feedback linearization backstepping control \cite{feedbacklinearization},
and model predictive control \cite{flatnessbasedMPC}. Among the numerous methods for quadcopter control, it is still rare to find a comprehensive control approach in the literature that offers four highly desired features: (i) anticipating future reference information, (ii) handling state and input
constraints, (iii) real-time implementability on on-board embedded hardware, and (iv) stability guarantees. Research into quadcopter trajectory tracking 
 control satisfying all these four features has mainly been pursued using Model Predictive Control (MPC) due to its ability to handle constraints, predictive behavior, and the availability of theoretical frameworks to provide stability guarantees
(see \cite{Alexpaper}).

In \cite{generalmpc-20l}, the differential flatness property of quadcopters is utilized to develop a real-time trajectory generation method. Conservative box constraints are then applied to create a convex feasible acceleration set using decoupled per-axis acceleration limits, followed by solving a convex optimization problem with bounds on each decoupled axis. In \cite{hierarchicalMPC-21}, a two-layer MPC approach is proposed, in which stabilization is managed by a linear constrained MPC controller, while at a higher hierarchical level and at a lower sampling rate, a hybrid MPC controller generates the online path for reaching a target position, considering obstacle avoidance. This real-time trajectory generation is especially advantageous in scenarios, where target and obstacle positions are unknown in advance, distinguishing it from offline planning methods. In \cite{feedbacklinearization}, a feedback linearization model predictive control algorithm is developed for quadcopter trajectory tracking, aiming to achieve a more accurate linearized predictive model compared to Taylor series-based approximations. However, the last cited three papers do not include formal stability guarantees. 

In addition to the linear MPC-based approach mentioned above, more recent papers have employed Nonlinear Model Predictive Control (NMPC) due to the enhanced onboard computational power of most quadcopters. In \cite{NMPC-26}, an LQR controller with integral action is applied in the outer loop, while NMPC is used for attitude control in the inner loop. Despite introducing terminal constraints in the NMPC problem to stabilize the system, there is no global stability guarantee offered for the closed-loop system.
%An NMPC controller is employed in \cite{NMPC-27} for obstacle avoidance, accommodating arbitrary non-convex-shaped obstacles.
In \cite{stableNMPC}, a stability-guaranteed NMPC scheme is presented for quadcopters, determining the minimal stabilizing prediction horizon for asymptotic stability. This work proposes distinct running cost to achieve stability
without relying on terminal conditions. %resulting in a larger region of attraction compared to many other stability-based controllers. 
Although of interest, a formal proof of stability is missing due to the forward Euler discretization of the nonlinear model instead of an exact method.

A cascade control approach, recently proposed in \cite{Alexpaper}, combines linear MPC for outer-loop position control with nonlinear inner-loop attitude control. This method ensures uniform almost global asymptotic stability (UaGAS) while managing the quadcopter’s thrust constraint. However, its real-life performance falls short due to the conservatism introduced by translating nonlinear model constraints to linear ones and adding an auxiliary stabilizing constraint in the MPC. Hence, a quadcopter controller with the above-mentioned four features (i)--(iv) and additional high tracking performance is currently lacking.

Motivated by this gap in the literature, we propose a cascade control structure with a new MPC strategy in the outer loop, offering all four key features of an effective control approach, while also providing high-performance in the sense of being capable of tracking fast references with small error. In the outer loop, the translational system, including position and velocity dynamics, is modeled as a 12th-order linear problem. The MPC strategy is then applied to address the outer-loop problem. The non-zero total thrust constraint from the original nonlinear model is translated into constraints for the linear model. 
%To reduce conservatism compared to \cite{Alexpaper}, we use TV constraints rather than the time-invariant constraints applied in \cite{Alexpaper}. Moreover, since the MPC law is formulated in discrete time, we ensure that constraints are satisfied between sample times.
Our approach decouples both dynamics and constraints, resulting in three separate 4th-order linear MPC formulations for the $x$, $y$, and $z$ directions, each subject to time-varying (TV) decoupled linear constraints. This decoupling enhances real-time implementability and is thus important to warrant feature (iii) of a desirable controller. These three MPC controllers generate a twice-differentiable virtual acceleration, which is then used to compute the thrust control input. The desired virtual acceleration is subsequently converted into a desired attitude, tracked in the inner loop by the nonlinear attitude controller from \cite{ErijenNLcontroller}. In \cite{Alexpaper}, a stabilizing input constraint is added to ensure uniform global asymptotic stability (UGAS) of the MPC strategy, which introduces conservatism and results in
reduced performance. Since the outer-loop dynamics are controllable and marginally stable\footnote{All eigenvalues lie within the unit circle and those on the unit circle are simple.}, we use a particular cost function proposed in \cite{neutrallystable} to ensure UGAS of the outer loop  without compromising performance. 
Compared to \cite{Alexpaper}, we propose a more efficient outer-loop MPC strategy for quadcopters by: (i) imposing less conservative constraints through time-varying constraints, rather than time-invariant ones, while still ensuring inter-sample constraint satisfaction; (ii) enhancing performance by employing an appropriate terminal cost instead of adding conservatism through auxiliary stabilizing constraints. These innovations require new technical developments and are instrumental for guaranteeing performance, as evidenced by high-fidelity numerical case studies. % and experimental validations.

The outline of this paper is as follows: Section \ref{section 2 preliminary} introduces the preliminaries and Section \ref{Dynamic section} presents the quadcopter dynamics model and the problem statement. Section \ref{section IV Controller design} introduces the cascade configuration and defines the inner-loop and outer-loop problems. Section \ref{section NLC} details the inner-loop controllers, while Section \ref{section OUTERLOOP} describes the structure of the outer loop. The MPC strategy in the outer loop is proposed in Section \ref{MPC1 Section}. The stability of the continuous-time system and cascade trajectory tracking are demonstrated in Sections \ref{section Stability of the continuous system} and \ref{cascade section}, respectively. Section \ref{case studies section} presents  numerical case studies, underlying high-fidelity models, %Section \ref{experiment section} covers the experimental validation,
and Section \ref{Conclusion section} concludes the work.
%\section{Preliminaries}
\vspace{-0.1cm}
\section{Preliminaries} \label{section 2 preliminary}
\vspace{-0.2cm}
This section introduces the notation, lemmas and theorems used throughout the paper. We utilize $\mathbb{R}$, $\mathbb{R}_{>0}$, and $\mathbb{R}_{\geq 0}$ to denote the set of real numbers, the set of positive real numbers, and the set of non-negative numbers, respectively. Additionally, $\mathbb{R}^n$ represents the set of all $n$-dimensional vectors of real numbers. The set of natural numbers is denoted by $\mathbb{N} = \{1,2,3 \hdots\}$, and $\mathbb{N}_0 = \mathbb{N} \cup \{0\}$. 
The symbol $I$ is an identity matrix of appropriate dimensions.
The 
%Euclidean norm,or 
2-norm of a vector $v\in \mathbb{R}^n$ is $\|v\|=\sqrt{v^{\T}v}$.
%The Frobenius norm of an $m$-by-$n$ matrix $A$ is defined by $\|A\|_F =\sqrt{\sum_{i=1}^{m}\sum_{j=1}^{n} |a_{ij}|^2}=\sqrt{\mathrm{trace}(A^{\T}A)}$, where the trace of a matrix is the sum of diagonal entries.
A diagonal matrix with diagonal elements $a_1, a_2, \ldots, a_n$ is denoted as $\text{diag}(a_1, a_2, \ldots, a_n)$.
$SO(3)$ is the group of 3-by-3 orthonormal matrices that represent rotations about the origin in
%three-dimensional Euclidean space,
$\mathbb {R} ^{3}$.
%preserving both length and orientation.
This work considers UGAS, 
%UGES, 
and uniform local exponential stability (ULES), with their definitions provided in \cite{Khalil:2002}, and adopts UaGAS, which is UGAS except for a measure-zero set of initial conditions, as defined in \cite{ErijenNLcontroller}. The saturation function $\mathrm{sat}:\mathbb{R}^m\to\mathbb{R}^m$ is defined as $ \mathrm{sat}(u)= [\mathrm{sat}(u_1),~ \mathrm{sat}(u_2),\hdots,~ \mathrm{sat}(u_m)]^{\T},$
\iffalse
\begin{equation*}
     \mathrm{sat}(u)= [\mathrm{sat}(u_1),~ \mathrm{sat}(u_2),\hdots,~ \mathrm{sat}(u_m)]^{\T},
\end{equation*}
\fi
where
\vspace{-0.2 cm}
\begin{equation}
\label{sat function}
{\mathrm{sat}(u_i)} = \begin{cases}
u_{\mathrm{max}},&\ u_i>u_{\mathrm{max}}\\ 
u_i,&\ |u_i|\leq u_{\mathrm{max}} \\
-u_{\mathrm{max}},&\ u_i<-u_{\mathrm{max}} 
\end{cases}
\end{equation}
for $i\in\{1,2,...,m\}$, assuming $u_{\mathrm{max}}>0$ is clear from the context. Consider the discrete-time system
\vspace{-0.2 cm}
\begin{equation}
    x^+=Ax+B\mathrm{sat}(u),
    \label{saturated system}
\end{equation}
where $x\in\mathbb{R}^n$ and $u\in\mathbb{R}^m$ denote the  current states and inputs, respectively. It is assumed that $(A,B)$ is controllable, and the system is marginally stable. In this paper, the stability proof of the proposed MPC strategy is built on the theorems and lemmas from \cite{neutrallystable}, which are repeated here for self-containment.
%According to \cite{neutrallystable}, 
\iffalse
For any $L_u$ satisfying $L_u u_{\mathrm{max}}>1$, 
\begin{equation*}
    \|\mathrm{sat}(u)-u\|\leq L_u u^{\T} \mathrm{sat}(u).
\end{equation*}
\fi

Marginal stability implies the existence of a positive definite matrix $M_c$, such that
\begin{equation}
    A^{\T} M_c A - M_c \preceq 0.
    \label{Mc}
\end{equation}
A globally stabilizing small-gain control is defined as
\begin{equation}
    u(x)=-\kappa B^{\T} M_c A x =: Kx,
    \label{smallgain}
\end{equation}
where $\kappa>0$ is chosen such that
\begin{equation}
    \kappa B^{\T} M_c B \prec I.
    \label{kappa}
\end{equation}
As demonstrated in \cite{neutrallystable}, there exists a positive definite matrix $M_q$ such that
\begin{equation}
    (A+BK)^{\T} M_q (A+BK) - M_q = -I.
    \label{Mq}
\end{equation}
The following theorem outlines the stability properties of the resulting closed-loop system.
\begin{thm} (\hspace{-0.003cm}\cite{neutrallystable}).
    For the closed-loop system \eqref{saturated system} with the small-gain control \eqref{smallgain}, assuming that $(A,B)$ is controllable and \eqref{saturated system} is marginally stable, there exist a global Lyapunov function $W: \mathbb{R}^n \rightarrow \mathbb{R}_{>0}$ such that 
\begin{align}
&W(x)=W_q(x)+\lambda W_c(x)= x^{\T} M_q x+\lambda(x^{\T} M_c x)^{3/2} \label{LyapFunc}\\
&W(Ax+B\mathrm{sat}(Kx))-W(x)\leq -\|x\|^2, 
\end{align}
with K in \eqref{smallgain}, $M_c \succ0$ as in (\ref{Mc}), $M_q\succ 0$ specified in (\ref{Mq}), and 
    \begin{equation}
        \lambda=\frac{2 \kappa L_u \sigma_{\mathrm{max}}(A_d^{\T} M_q B_d)}{\sqrt{\lambda_{\mathrm{min}}(M_c)}},
        \label{lambda eq}
    \end{equation}
    where $\sigma_{\mathrm{max}}$ and $\lambda_{\mathrm{min}}$ denote the maximum singular value and minimum eigenvalue of a matrix, respectively and $L_u$ is chosen such that $L_u u_{\mathrm{max}}>1$.
\label{Thoery small gain}
\end{thm}
 On the basis of the Theorem \ref{Thoery small gain} and using the following lemma, a globally stabilizing MPC for (\ref{saturated system}) can be derived.
\begin{lem} (\hspace{-0.002cm}\cite{neutrallystable}).
\label{lemma 1}
    Consider a discrete-time system described by $x^+=Ax+Bu,$ where $x\in \mathbb{R}^n$, $u\in \mathbb{R}^m$, $(A,B)$ is controllable, and the optimal control sequence $u^*(x)=\big(u^*_{0|k}(x), u^*_{1|k}(x),...,u^*_{N|k}(x)\big)$ obtained from
\begin{align}
\label{general MPC problem}
    \min_{u}~&~ J_N(x,u) = V(x_{N|k})+\sum_{i=0}^{N-1}l(x_{i|k},u_{i|k})\\
    \mathrm{s.t.}~&~  x_{0|k}= x,\nonumber\\
    &~x_{i+1|k}= A x_{i|k}+B u_{i|k},~i\in \{0,1,..,N-1\},\nonumber\\
    &~u_{i|k} \in  \mathscr{U},~i\in \{0,1,..,N-1\},\nonumber
\end{align}
where 
\begin{equation}
\label{l}
\hspace*{-1cm} % Adjust the value as needed
\begin{aligned}
    & l(x_{i|k},u_{i|k})=x_{i|k}^{\T}Qx_{i|k}+u_{i|k}^{\T}R ~u_{i|k},
\end{aligned}
\end{equation}
with $Q$ and $R$ positive definite,  $V:\mathbb{R}^n  \rightarrow \mathbb{R}_{>0}$ represents the terminal cost and is chosen such that $V$ is a control Lyapunov function, $\mathscr{U}$ represents the input constraint set, and $J_N^*(x)=J_N(x,u^*)$ is the optimal cost.
If there exist a local control law $k_f:\mathbb{R}^n  \rightarrow \mathbb{R}^m$ such that
\begin{enumerate}
    \item $k_f(x) \in \mathscr{U}$;
    \item $V(Ax+Bk_f(x))-V(x)+l(x,k_f(x))\leq 0$;
    \item $\alpha_1(\|x\|)\leq J^*_N(x)\leq \alpha_2(\|x\|)$,
\end{enumerate}
with $\alpha_1$ and $\alpha_2$ $\mathscr{K}_{\infty}$ -functions\footnote{A function $ \alpha : [0, \infty) \to [0, \infty)$ is in class $\mathscr{K}_{\infty}$ if it is strictly increasing, $\alpha(0)=0$, and $\alpha(r)\to \infty$ as $r \to \infty$ (\hspace{-0.0003 cm}\cite{neutrallystable}).},
then the MPC optimization problem is guaranteed feasible, guaranteeing $x^+=Ax+Bk_N(x)$ is globally asymptotically stable, and
%In addition, $J^*_N(x)$ is monotonically non-increasing such that
\begin{equation*}
    J^*_N(x^+)=J^*_N(Ax+Bk_N(x))\leq J^*_N(x)-l(x,k_N(x)),
\end{equation*}
with the MPC law
\vspace{-0.3cm}
\begin{equation}
    k_N(x)=u^*_{0|k}(x).
    \label{kn}
\end{equation}
\vspace{-0.6cm}
\end{lem}
A globally stabilizing MPC strategy for the system \eqref{saturated system} can now be derived, as outlined in Theorem \ref{GAS theorem}.
\begin{thm}  
\label{GAS theorem}
(\! \cite{neutrallystable}).
    Consider the closed-loop system \eqref{saturated system}, with the control law $k_N(x)$ as in (\ref{kn}), resulting from the optimization problem
    \vspace{-0.05cm}
%\begin{subequations}
\begin{align}
    \min_{u} ~&~ J_N(x,u) = V(x_{N|k})+\sum_{i=0}^{N-1}l(x_{i|k},u_{i|k}) \label{stable MPC problem} \\
    \mathrm{s.t.}~&~  x_{0|k}= x, \nonumber\\
    &~x_{i+1|k}= A x_{i|k}+B u_{i|k},~i\in \{0,1,..,N-1\}, \nonumber \\
    &~u_{i|k} \in [-u_{\mathrm{max}}~,~u_{\mathrm{max}}],~i\in \{0,1,..,N-1\}, \nonumber
\end{align}
%\end{subequations}

where $V(x_{N|k})=\Theta W(x_{N|k})$, $W$ as in \eqref{LyapFunc}, $l$ defined in (\ref{l}), $N \in \mathbb{N}$ is the prediction horizon, $Q$ and $R$ being 
%arbitrary positive definite matrices,
positive definite and  $\Theta$ is a positive constant such that
\begin{equation}
    \Theta\geq \lambda_{max}(Q+\kappa^2 A^{\T} M_c B R B^{\T} M_c A).
    \label{theta}
\end{equation}
Then, given any positive integer $N$, the closed-loop system $x^+=Ax+Bk_N(x)$ is globally asymptotic stable and 
\begin{equation}
    J_N^*(Ax+Bk_N(x))-J_N^*(x)\leq -l(x,k_N(x))
    \label{Jstar NE}
\end{equation}
for all $x\in \mathbb{R}^n$.
\end{thm}
\begin{rem}
 Given that for time-invariant finite-dimensional systems, global asymptotic stability is equivalent to UGAS, we can infer that the MPC control strategy outlined in \mbox{Theorem \ref{GAS theorem}} ensures UGAS.
%—where the decay rate and transient overshoot of solutions are uniform on bounded sets of initial states—
\end{rem} 
Due to the cubic term in the terminal cost, the proposed MPC in Theorem \ref{GAS theorem} is not a QP problem, but it is convex and is solvable with convex optimization solvers.
\vspace{-0.2cm}
\section{Dynamics and Problem Setup} \label{Dynamic section}
\vspace{-0.2cm}
The model used in this work is based on the one presented in \cite{Alexpaper}. Let $G$ be a right-handed inertial (or world) frame using the North-East-Down (NED) convention, with $\{x_G, y_G, z_G\}$ as its orthonormal basis vectors. $B$ represents a right-handed body-fixed frame with orthonormal basis vectors $\{x_B, y_B, z_B\}$, representing the axes of $B$ relative to $G$. The rotation matrix  $R = [x_B, y_B, z_B] \in SO(3)$ defines the orientation of the body-fixed frame relative to the world frame. In addition, angular velocity, linear velocity and position of the body with respect to the world frame are described with vectors $\omega = [\omega_1, \omega_2, \omega_3]^{\T}$, $v = [v_x, v_y, v_z]^{\T}$ and $p = [p_x, p_y, p_z]^{\T} \in \mathbb{R}^3$, respectively.
With the defined variables, the dynamics of the quadcopter can be described as
\vspace{-0.15 cm}
\begin{subequations}\label{dynamics}
\begin{align}
 \Dot{p}&=v, \label{position}\\
\Dot{v}&= g z_G-T z_B - D^{\T}v, \label{velocity}\\
\Dot{R}&=RS(\omega), \label{Rotation}\\
J\Dot{\omega}&=S(J\omega)\omega-\tau_g-AR^{\T}v-C\omega+\tau, \label{angularvalocity}
\end{align}
\end{subequations}
where $g$ is the gravitational acceleration, $T \geq 0$ represents the magnitude of the combined thrust of the four propellers, normalized by mass, $D = \operatorname{diag}(d_x, d_y, d_z)$, $d_x, d_y, d_z> 0$, are the mass-normalized rotor drag coefficients, $\tau_g \in \mathbb{R}^3$ are torques resulting from gyroscopic effects, $J \in \mathbb{R}^{3 \times 3}$ is the inertia matrix, $A$ and $C$ are constant matrices, \mbox{$\tau = [\tau_1, \tau_2, \tau_3]^{\T} \in \mathbb{R}^3 $} is the torque input and $S(\omega)$ represents a skew-symmetric matrix such that $S(a)b=a \times b$ for any vectors $a,b \in \mathbb{R}^3$.
\iffalse
\begin{equation*}
S(\omega)=-S(\omega)^{\T}=
\begin{bmatrix} 
0 & -\omega_3 & \omega_2\\
\omega_3 & 0 & -\omega_1 \\
-\omega_2 & \omega_1& 0
 \end{bmatrix}.
\end{equation*}
Note that a linear drag model is considered, similar to \cite{Alexpaper}.
Furthermore, it is assumed that the effects of rotation on the drag force are negligible, implying that the drag force can be approximated as $RDR^{\T}\approx D$ in (\ref{velocity}).
\fi

The thrust is non-negative and limited according to
\begin{equation}
    0 \leq T(t) \leq T_{\mathrm{max}},~~ \text{for~all~} t \in \mathbb{R}_{\geq 0},
    \label{constraint}
\end{equation}
where $T_{\mathrm{max}} > g$. This physical restriction arises because the propellers can generate only a limited upward thrust and must counteract the gravitational force.
\iffalse
\vspace{-0.1cm}
\subsection{Reference trajectory and problem statement}
\vspace{-0.2cm}
\fi

A reference trajectory $\big(\bar{p}, \bar{v}, \bar{R}, \bar{\omega}, \bar{T}(t), \bar{\tau}\big) : \mathbb{R}_{\geq 0} \rightarrow \mathbb{R}^3 \times \mathbb{R}^3 \times SO(3) \times \mathbb{R}^3 \times \mathbb{R} \times \mathbb{R}^3 $ is referred to be a feasible trajectory, if it satisfies the dynamics (\ref{dynamics}) for all $t \in \mathbb{R}_{\geq 0}$, i.e.,
\begin{subequations}\label{reference}
\begin{align}
\Dot{\bar{p}} &= \bar{v}, 
    & \Dot{\bar{v}} &= g z_G - \bar{T} \bar{z}_B - D\bar{v}, \label{pv ref} \\
\Dot{\bar{R}} &= \bar{R}S(\bar{\omega}), 
    & J\Dot{\bar{\omega}} &= S(J\bar{\omega})\bar{\omega} - \tau_g - A\bar{R}^{\T}\bar{v} - C\bar{\omega} + \bar{\tau}, \label{RW ref}
\end{align}

\end{subequations}
and $0 < \epsilon_1 \leq \bar{T}(t) \leq T_{\mathrm{max}}-\epsilon_2,$
with fixed $\epsilon_1, \epsilon_2>0$.

For a feasible reference trajectory, the error coordinates are defined as
\begin{equation}
 \label{error dynamics}
 \Tilde{p}=\bar{p}-p,~~ \Tilde{v}=\bar{v}-v,~~\Tilde{R}=\bar{R}^{\T} R,~~
 \Tilde{\omega}= \omega-\Tilde{R}^{\T} \bar{\omega}.
\end{equation}
The main tracking control problem of the quadcopter can now be formulated as follows.
\begin{problem}
\label{problem1}
    (\hspace{-0.0003 cm}\cite{Alexpaper}). Given a feasible reference trajectory $(\bar{p}, \bar{v}, \bar{R}, \bar{\omega}, \bar{T}, \bar{\tau})$, develop appropriate control laws
\begin{align*}
    T=T(p,v, R, \omega, \bar{p}, \bar{v}, \bar{R}, \bar{\omega}),~~
    \tau=\tau(p,v, R, \omega, \bar{p}, \bar{v}, \bar{R}, \bar{\omega}),
\end{align*}
such that (\ref{constraint}) holds and  the origin $(\Tilde{p}, \Tilde{v}, \Tilde{R}, \Tilde{\omega})=(0, 0, I, 0)$ of the resulting closed-loop system is UaGAS.
\end{problem}

\section{Controller Design} \label{section IV Controller design}
\begin{figure}
[!t]\centering 
\includegraphics[width=0.45\textwidth]{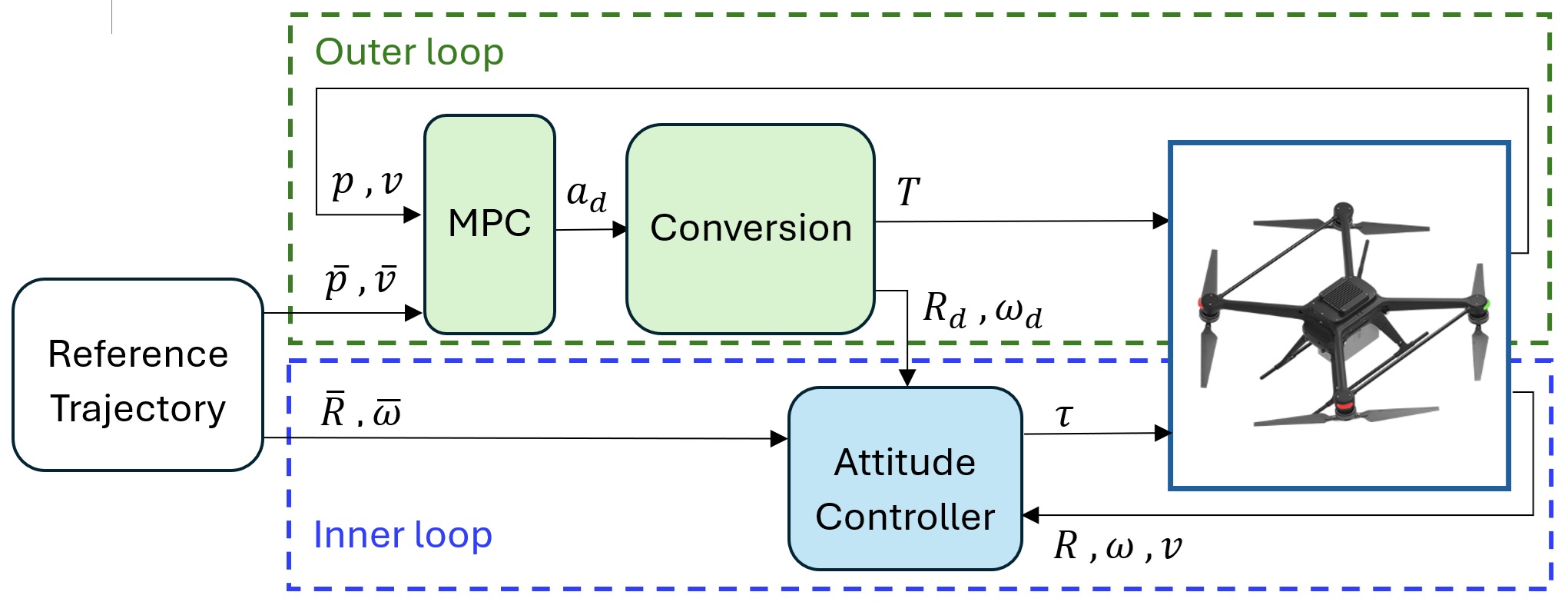}
\caption{Overview of the inner–outer loop control scheme: The outer loop MPC generates desired acceleration $a_d$, which is converted to thrust $T$, desired attitudes $R_d$, and angular velocities $\omega_d$, with $T$ applied to the quadcopter. The inner-loop attitude controller uses $R_d$, $\omega_d$, and measured attitudes and angular velocities to generate torque inputs $\tau$ for the quadcopter.}
\label{config}
\end{figure}

\vspace{-0.1cm}
 To address the tracking control problem outlined in Problem \ref{problem1}, a cascade control architecture is employed, consisting of an outer-loop and an inner-loop problem, that contain the translational \eqref{position}, \eqref{velocity} and rotational \eqref{Rotation}, \eqref{angularvalocity} dynamics, respectively, see Fig.~\ref{config}. This section specifies the cascade configuration and introduces the problem definitions for the inner-loop and outer-loop problems.
\vspace{-0.2cm}
\subsection{Cascade trajectory tracking setup and constraints}
\label{section 4.1}
\vspace{-0.2cm}
The dynamics for position and velocity errors in (\ref{error dynamics}) follow from  (\ref{pv ref}), (\ref{position}), and (\ref{velocity}), are given by 
\begin{subequations}
\begin{align*}
\Dot{\Tilde{p}}=\Tilde{v},~
\Dot{\Tilde{v}}=-D\Tilde{v}+Tz_B-\bar{T}\bar{z}_B.
\end{align*}
\end{subequations}
A desired acceleration error $a_d \in \mathbb{R}^3$ is defined as
\begin{equation}
    a_d=Tz_B-\bar{T}\bar{z}_B.
    \label{ad virtual}
\end{equation}
So, $T$ and $z_B$ have to be chosen such that \eqref{ad virtual} holds. This allows the translational system, including the position and velocity dynamics, to be reformulated as a linear system
\begin{equation}
   \label{outerloop dynamic} 
\Dot{\Tilde{p}}=\Tilde{v},~
\Dot{\Tilde{v}}=-D\Tilde{v}+a_d.  
\end{equation}
%In Section \ref{section OUTERLOOP}, an MPC strategy is employed to generate $a_d$.
%the desired virtual acceleration

%, ensuring both the stabilization of the dynamics in (\ref{outerloop dynamic}) and satisfaction of the constraint in (\ref{constraint}). 
Thus, the first input $T$ can be determined by
\begin{equation} \label{thrust}
    T=\|a_d+\bar{T}\bar{z}_B\|.
\end{equation}
To track the desired virtual acceleration generated by the MPC in the outer loop, a desired thrust vector is generated and converted into a desired attitude. This attitude is then tracked using the torque $\tau$ generated by the attitude tracking controller presented in Section \ref{section NLC}. In \cite{Alexpaper}, it is demonstrated how to achieve the desired attitude using the expression
\begin{equation} \label{Rd}
    R_d=[x_{B,d},~ y_{B,d},~ z_{B,d} ],
\end{equation}
where
\begin{align*}
z_{B,d}=&\begin{bmatrix} 
z_{B,d_1}&
z_{B,d_2}&
z_{B,d_3}
 \end{bmatrix}^{\T}=\frac{\bar{R}^{\T} a_d + \bar{T}e_3}{\|\bar{R}^{\T} a_d + \bar{T}e_3\|}, \\
y_{B,d}=&\begin{bmatrix} 
0&
\frac{z_{B,d_3}}{\sqrt{z_{B,d_2}^2+z_{B,d_3}^2}} &
\frac{-z_{B,d_2}}{\sqrt{z_{B,d_2}^2+z_{B,d_3}^2}}
 \end{bmatrix}^{\T},\nonumber\\
 x_{B,d}=&\begin{bmatrix} 
\sqrt{z_{B,d_2}^2+z_{B,d_3}^2}&
\frac{-z_{B,d_1}z_{B,d_2}}{\sqrt{z_{B,d_2}^2+z_{B,d_3}^2}} &
\frac{-z_{B,d_1}z_{B,d_3}}{\sqrt{z_{B,d_2}^2+z_{B,d_3}^2}}
 \end{bmatrix}^{\T}.\nonumber
\end{align*}
Subsequently, applying $\Dot{R_d}=R_d S(\omega_d)$ yields an expression for the desired angular velocity \cite{Alexpaper}
\begin{equation} \label{wd}
    \omega_d=\begin{bmatrix} 
-y_{B,d}^{\T} \Dot{z}_{B,d}&
\hphantom{-}x_{B,d}^{\T} \Dot{z}_{B,d} &
~-x_{B,d}^{\T} \Dot{y}_{B,d}
 \end{bmatrix}^{\T}.
\end{equation}
These derivations indicate that $a_d$ can be tracked by determining the desired attitude in (\ref{Rd}) and applying the thrust specified in (\ref{thrust}). This means that when the errors (\ref{outerloop dynamic}) converge to zero, $a_d$
  also converges to zero and $z_{B,d}\rightarrow e_3$, causing $R_d \rightarrow I$. The inner-loop controller requires setpoints for the desired angular velocity and its derivative, so the desired acceleration must be twice differentiable. 
  
%According to (\ref{constraint}), the magnitude of the thrust vector is constrained. For $z_{B,d}$ and $y_{B,d}$  to be well-defined, $T\neq0$ must hold, and $z_{B,d}$ and $e_1$ should never be parallel, achieved by requiring $z_{B,d_3}>0$. 
The constraint (\ref{constraint}) can be converted into a constraint on $a_d$ by considering the following set of admissible values for the desired acceleration \cite{Alexpaper}:
\begin{multline}
\label{main ad constraint}
\mathcal{A}(\bar{R}, \bar{T})=\{a_d \in \mathbb{R}^3 ~|~ 0 <  \| a_d+\bar{T} \bar{z}_B  \| \leq T_{\mathrm{max}},\\
\bar{z}_B^{\T} a_d+\bar{T}>0 \}.
\end{multline}
\vspace{-0.85cm}
\subsection{Cascade Problem Definition}
\vspace{-0.2cm}
To address Problem \ref{problem1} using a cascade approach, we define two sub-problems: The outer-loop and inner-loop problems.
\vspace{-0.1cm}
\begin{problem}[Outer-loop problem]
\label{outerloop problem}
Find a virtual acceleration control law $a_d = a_d(p,v,\bar{p}, \bar{v}, \bar{T})$, which is twice differentiable with respect to time, such that the origin $(\Tilde{p}, \Tilde{v})=(0,0)$ of the system (\ref{outerloop dynamic}) is UGAS and $a_d(t) \in \mathcal{A}(\bar{R}, \bar{T})$ for all $t \in \mathbb{R}_{\geq 0}$, with ${A}(\bar{R}, \bar{T})$ as defined in \eqref{main ad constraint}.
\end{problem}
Given the goal of steering $\Tilde{R}$ to $R_d$ in the inner loop, attitude and angular velocity errors are formulated as 
\begin{subequations}\label{Re-we}
\begin{align}
R_e&= R_d ^{\T} \Tilde{R}, \label{Re} \\
\omega_e&=\omega-\Tilde{R}^{\T} \bar{\omega}-R_e^{\T} \omega_d. \label{we} 
\end{align}
\end{subequations}
The inner-loop problem is now defined as follows:
\begin{problem}[Inner-loop problem]
\label{innerloop problem}
Find a control law $\tau = \tau(R,\omega,\bar{R}, \bar{\omega}, \bar{T}, a_d)$, such that the origin $(R_e,\omega_e)=(I,0)$ of the system (\ref{Re-we}) is UaGAS.
\end{problem}
\vspace{-0.1cm}
%In \cite{Alexpaper} it is demonstrated that solving Problems 2 and 3 yields a solution to Problem 1.
%\vspace{-0.1cm}
\section{Inner-loop Tracking} \label{section NLC}
\vspace{-0.25cm}
In this work, the non-linear controller from \cite{ErijenNLcontroller} is employed for stabilizing the attitude dynamics of the quadcopter, as it ensures ULES and UaGAS for the attitude dynamics.
\iffalse
The error variables in (\ref{Re-we}) are described by
\begin{align*}
\Dot{R_e}&=R_e S(\omega_e), \\
J\Dot{\omega}_e&=S(J\omega)\omega - \tau_g -AR^{\T} v -C \omega + \tau \\
& -J\Tilde{R}^{\T} J^{-1}(S(J\bar{\omega})\bar{\omega}-\tau_g - A\bar{R}^{\T} \bar{v} - C \bar{\omega}+\bar{\tau}) \\
&+J[(S(\omega)\Tilde{R}^{\T}-\Tilde{R}^{\T} S(\bar{\omega}))\bar{\omega}+S(\omega_e)R_e^{\T}\omega_d-R_e^{\T} \Dot{\omega_d}], 
\end{align*}
\fi
The input is given by
\begin{align}
\label{tau}
\begin{split}
\tau&=-K_{\omega}\omega_e + K_R \sum_{i=1}^{3} k_i(e_i \times R_e^{\T} e_i)\\
&-S(J\omega)\omega + \tau_g +AR^{\T} v +C \omega \\
&+ J\Tilde{R}^{\T} J^{-1}(S(J\bar{\omega})\bar{\omega}-\tau_g - A\bar{R}^{\T} \bar{v} - C \bar{\omega}+\bar{\tau}) \\
&-J[(S(\omega)\Tilde{R}^{\T}-\Tilde{R}^{\T} S(\bar{\omega}))\bar{\omega}+S(\omega_e)R_e^{\T}\omega_d-R_e^{\T} \Dot{\omega_d}],
\end{split}
\end{align}
which results in the closed-loop system
\begin{subequations}
\begin{align*}
\Dot{R_e}=R_e S(\omega_e),~~
J\Dot{\omega}_e= -K_{\omega}\omega_e + K_R \sum_{i=1}^{3} k_i(e_i \times R_e^{\T} e_i), 
\end{align*}
\end{subequations}
for some $k_i>0$, $K_{\omega}>0$ and $K_R>0$. Since \( R_e \) converges to \( I \) for almost all initial conditions, it can be concluded from (\ref{Re}) that \( \Tilde{R} \) converges to \( R_d \) for almost all initial conditions. Additionally, as \( \omega_e \rightarrow 0 \), (\ref{we}) implies that \( \Tilde{\omega} \rightarrow R_e^{\T} \omega_d \). Combining this with \( R_e \rightarrow I \) results in \( \Tilde{\omega} \rightarrow \omega_d \), thus solving Problem \ref{innerloop problem}, see  \cite{Alexpaper} for more details.
\vspace{-0.4cm}
\section{Outer-loop Dynamics and Constraints} \label{section OUTERLOOP}
\vspace{-0.2cm}
\iffalse
To address the outer-loop problem outlined in Problem \ref{outerloop problem}, a model predictive control (MPC) approach is adopted. 
\subsection{Discretization and input transformation}
\fi
As the MPC law employed in the outer loop will be formulated in discrete time, to ensure that the desired accelerations $a_d$
  remain twice differentiable, the translational error dynamics in (\ref{outerloop dynamic}) are extended to a 12th-order linear model \cite{Alexpaper} 
  \vspace{-0.2 cm}
\begin{subequations}\label{Main outerloop dynamics}
\begin{align}
\Dot{\Tilde{p}}&=\Tilde{v}, \label{position error dynamic} \\
\Dot{\Tilde{v}}&=-D\Tilde{v}+a_d,\label{velocity error dynamic} \\
 \Dot{a}_d&=-\frac{1}{\gamma}(a_d-\eta),
 \label{ad dynamic} \\
\Dot{\eta}&=-\frac{1}{\gamma}(\eta-s), \label{eta dynamic}
\vspace{-0.2cm}
\end{align}
\end{subequations}
with $\gamma>0$,  $\Tilde{p}, \Tilde{v}, a_d, \eta \in \mathbb{R}^3$ are the states and $s \in \mathbb{R}^3$ is the input. 
Note that the dynamics in (\ref{Main outerloop dynamics}) are decoupled, allowing the dynamics in the \(x\), \(y\), and \(z\) directions to be separated into three distinct fourth-order systems.

\iffalse
A discrete-time MPC law is formulated based on exact discretization of (\ref{Main outerloop dynamics}), using zero-order hold (ZOH),
\begin{equation*}
    s(t)=s(t_k),~~t\in[t_k,t_{k+1}),
\end{equation*}
with $t_k=kh$, $k \in \mathbb{N}_0$, and $h>0$ representing the sampling period. This yields the discrete-time system
\begin{equation}
    x[k+1]=A_d x[k]+ B_d u[k],
    \label{DT system dynamics coupled}
\end{equation}
where $x[k]=x(t_k)=[\Tilde{p}(k) ~ \Tilde{v}(k)~ a_{d}(k)~ \eta(k)]^{\T}\in\mathbb{R}^{12}$ represents the system states, $u(k)=s(t_k)=[u_1(k)~ u_2(k)~u_3(k)] \in \mathbb{R}^3$ denotes the system input, and  $A_d$ and $B_d$ represent the discrete-time state-space matrices obtained by zero-order hold discretization of the continuous-time dynamic system model (\ref{Main outerloop dynamics}).
\subsection{Constraints}
\label{constraint section}
\fi

\begin{figure}
\centering
\includegraphics[width=1.9in]{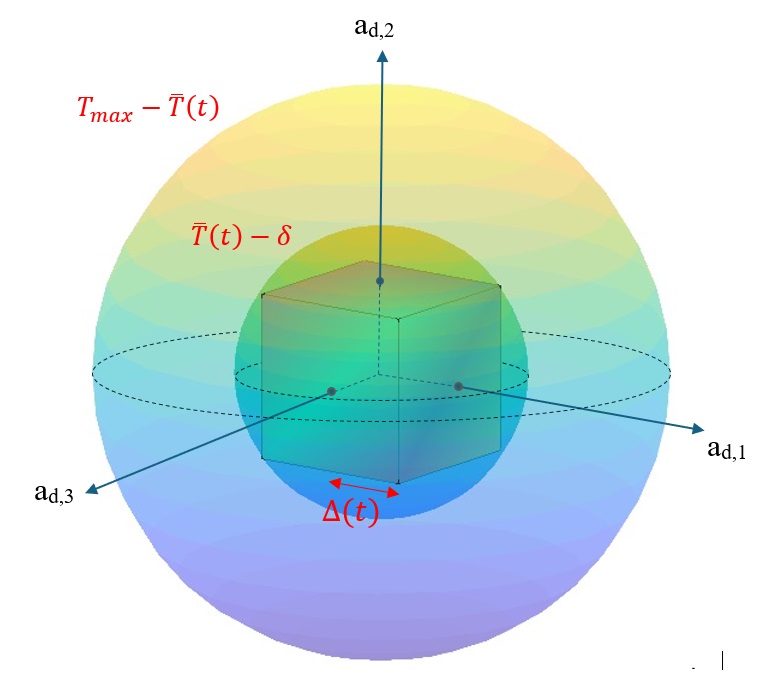}
\caption{Schematic of TV constraints on \( a_d \) in 3D. The radius of each sphere varies with the reference thrust \( \bar{T} \), so the sphere with radius \( \bar{T}(t)-\delta \) is not always the smaller one. The cube illustrates the decoupled constraint, limiting each of the components of $a_d$ to be within $\pm \Delta(t)$.}
\label{sperecube}
\end{figure}

As detailed in Section \ref{section 4.1}, $a_d$ must lie in the admissible set defined in (\ref{main ad constraint}). Given that $\| a_d+\bar{T} \bar{z}_B  \| \leq \| a_d\|+\bar{T} $, when $\| a_d\|+\bar{T}\leq T_{\mathrm{max}}$,
the condition $\| a_d+\bar{T} \bar{z}_B  \| \leq T_{\mathrm{max}}$ in (\ref{main ad constraint}) is satisfied, albeit with some conservatism. The second condition in (\ref{main ad constraint}), i.e., $\bar{z}_B^{\T} a_d+\bar{T}>0$ is met, if $\|a_d\|\leq \Bar{T}-\delta,$
for some small $0<\delta<\epsilon_1$. Therefore, a new and smaller set of admissible desired acceleration values, $\mathcal{A}_o(\bar{T}) \subset \mathcal{A}(\bar{R}, \bar{T})$, can be defined as 
\vspace{-0.15 cm}
\begin{equation}
\mathcal{A}_o(\bar{T})=\{a_d \in \mathbb{R}^3 ~|~  \| a_d \| \leq \rho (t)\},
   \label{TV cons coupled}
\end{equation}
where
\vspace{-0.4 cm}
\begin{equation}
\label{R(t)}
    \rho (t)=\min(\bar{T}(t)-\delta~,~T_{\mathrm{max}}-\bar{T}(t)).
\end{equation}
The approach in this work involves decoupling the nonlinear constraint in \eqref{TV cons coupled}. A box approximation is used, representing the largest cube that fits within the smaller sphere at each time instance. While this simplifies implementation, it introduces some level of conservatism. See Fig.~\ref{sperecube}.
In this scenario, $a_d$ is constrained to lie in a more conservative set, $\mathcal{A}_c(\bar{T}) \subset \mathcal{A}_o(\bar{T}) \subset \mathcal{A}(\bar{R}, \bar{T})$, such that
\begin{equation}
\mathcal{A}_c(\bar{R},\bar{T})=\{a_d \in \mathbb{R}^3 \big|~ |a_{d,i}(t)| \leq \Delta(t),i\in\{1,2,3\}\},
   \label{TV cons decoupled}
\end{equation}
where $\Delta(t)=\frac{1}{\sqrt{3}}\rho (t)$.

Decoupling the constraints simplifies control design and reduces online computentional effort by solving three smaller 4th-order problems with decoupled linear constraints, but introduces conservatism due to the decoupled constraints.
Note that in \cite{Alexpaper}, the constraint is made time-invariant by considering the smallest sphere over all time, lower-bounding the TV bound in \eqref{TV cons decoupled} by \( \Delta = \inf_{t \in \mathbb{R}_{\geq 0}} \Delta(t) \), which introduces additional conservatism.
To reduce conservatism in constraints compared to %the approach outlined in 
\cite{Alexpaper}, we propose three separate linear MPC formulations, each subject to TV decoupled linear constraints instead of the time-invariant constraints. The TV nature of the constraints requires a significantly different MPC design compared to [2], as we will see in Section \ref{MPC1 Section}. In our approach, $\mathcal{A}_o(\bar{T})$ is approximated by a TV cube with a side length of $\frac{2\rho(t)}{\sqrt{3}}$. See Fig.~\ref{sperecube}.
\vspace{-0.25cm}
\section{MPC Formulation} \label{MPC1 Section}
\vspace{-0.3cm}
Upon decoupling the system (\ref{Main outerloop dynamics}), the dynamics of $x, y,$ and
$z$ are separated into three linear fourth-order models:
\begin{equation}
\begin{aligned}
  {\Dot{\Tilde{p}}}^i &= \Tilde{v}^i, & \Dot{\Tilde{v}}^i &= -d^i \Tilde{v}^i + a_d^i, \\
  \Dot{a}_d^i &= -\frac{1}{\gamma}(a_d^i - \eta^i), & \Dot{\eta}^i &= -\frac{1}{\gamma}(\eta^i - s^i),
\end{aligned}
\label{decoupled outerloop dynamics}
\end{equation}
where \(i \in \{1,2,3\}\), \(\gamma > 0\), \(\Tilde{p}^i, \Tilde{v}^i, a_d^i, \eta^i \in \mathbb{R}\) are the states and \(s^i \in \mathbb{R}\) is the input. The system is subject to the TV constraint \(  |a_{d,i}(t)| \leq \Delta(t) \), $t \in \mathbb{R}_\geq0$, as in (\ref{TV cons decoupled}). The MPC law will be formulated in discrete time, thereby requiring the conversion of (\ref{decoupled outerloop dynamics}) to a discrete-time model. Using exact zero-order hold (ZOH) discretization, with $s(t)=s(t_k),~t\in[t_k,t_{k+1})$, $t_k=kh$, $k \in \mathbb{N}_0$, and $h>0$ representing the sampling period, leads to the discrete-time system
\begin{equation}
    x^+=A_d x+ B_d u,
    \label{DT system dynamics}
\end{equation}
where $x=x(t_k)$, $x=[\Tilde{p}~ \Tilde{v}~ a_d~ \eta]^{\T} \in \mathbb{R}^4$ representing the system states, and $u=s(t_k) \in \mathbb{R}$ denotes the system input. The system matrices $A_d$ and $B_d$ are displayed in Table \ref{tab:ss matrices}. Note that $A_d$ has three eigenvalues inside the unit circle and one simple eigenvalue on the unit circle, indicating marginal stability of the system.
The main continuous-time problem is subject to TV state constraints defined in (\ref{TV cons decoupled}). However, to obtain a discrete-time MPC setup, this constraint must be transformed into specific constraints on the sample times, while ensuring inter-sample constraints satisfaction.
The following MPC problem is formulated for the outer loop:
\vspace{-0.3cm}
\begin{subequations}
\label{MPC problem 1}
 \begin{align}
    \min_{U_k}~&~ J(x_k,U_k) = V(x_{N|k})+\sum_{i=0}^{N-1}(x_{i|k}^{\T}Qx_{i|k}+u_{i|k}^{\T}R ~u_{i|k})\nonumber\\
   \mathrm{s.t.}~&~ x_{0|k}=~x_k,\nonumber\\
    &~x_{i+1|k}=~A_d x_{i|k}+B_d u_{i|k},~i\in \{0,1,..,N-1\},\nonumber\\
    &~|u_{i|k}| \leq~ \Delta_{i|k},~~i\in \{0,1,..,N-1\},\label{35a}\\
     &~|a_{d_{i|k}}|\leq~ \Delta_{i|k},~i\in \{0,1,..,N\},\label{35b}\\
     &~ |\eta_{i|k}|\leq~ \Delta_{i|k},~~i\in \{0,1,..,N\},\label{35c}
 \end{align}
 \end{subequations}
where
\begin{equation}
\label{Deltak definition}
\hspace*{-2.5cm} % Adjust the value as needed
    \begin{aligned}
    &\Delta_{i|k}:=\min_{t \in [t_{k+i}~,~t_{k+i+1}]} \Delta(t),
\end{aligned}
\end{equation}
with $\Delta(t)$ defined in (\ref{TV cons decoupled}), \mbox{$U_k=[u_{0|k}, u_{1|k}~\cdots~u_{N-1|k}]^{\T}$} contains the predicted future control inputs, $N \in \mathbb{N}$ is the prediction horizon, $V:\mathbb{R}^4 \rightarrow \mathbb{R}_{\geq 0}$ represents the terminal cost and is chosen such that $V$ is a control Lyapunov function, $Q$ and $R$ are positive definite, $x_k$ is the state at discrete-time step $k$, and $x_{i|k}$, $u_{i|k}$ denote the prediction of the state and input at time step $i+k$, respectively, based on information available at time $k$.

The following sections address inter-sample constraint satisfaction, the conversion of state to input constraints, and the UGAS guarantee of the proposed MPC law.

\begin{table*}
\caption{State space matrices of the system in (\ref{DT system dynamics}) \label{tab:ss matrices}}
\centering
\begin{tabular}{r l}
\hline
\vspace{0.1cm}
$A_d = $ &
$\begin{bmatrix}
    1 & \frac{1-e^{-dh}}{d} &
    \frac{\gamma (e^{-dh}+d\gamma-d\gamma e^{-\frac{h}{\gamma}}-1)}{d(d\gamma -1)} &
    \frac{\gamma (e^{-dh}+2d\gamma-d^2\gamma^2-2d\gamma e^{-\frac{h}{\gamma}}-dhe^{-\frac{h}{\gamma}}+d^2\gamma^2e^{-\frac{h}{\gamma}}+d^2\gamma he^{-\frac{h}{\gamma}} -1)}{d(d\gamma -1)^2} \\
    0 & e^{-dh} & \frac{\gamma(e^{-\frac{h}{\gamma}}-e^{-dh})}{d\gamma -1} &
    \frac{\gamma e^{-\frac{h}{\gamma}}+h e^{-\frac{h}{\gamma}}-\gamma e^{-dh}-d\gamma he^{-\frac{h}{\gamma}}}{(d\gamma -1)^2} \\
    0 & 0 & e^{-\frac{h}{\gamma}} & \frac{h}{\gamma}e^{-\frac{h}{\gamma}} \\
    0 & 0 & 0 & e^{-\frac{h}{\gamma}}
\end{bmatrix}$ \\
$B_d = $ &
$\begin{bmatrix}
    \frac{e^{-dh}+dh+3d^2\gamma^2-2d^3\gamma^3+d^3\gamma^2 h-3d^2 \gamma^2e^{-\frac{h}{\gamma}}+2d^3 \gamma^3e^{-\frac{h}{\gamma}}-2d^2\gamma h +d^3 \gamma^2 h e^{-\frac{h}{\gamma}}-d^2 \gamma h e^{-\frac{h}{\gamma}} -1}{d^2(d\gamma -1)^2} \\
    \frac{1-e^{-dh}-2d\gamma+d^2 \gamma^2+2d\gamma e^{-\frac{h}{\gamma}} +d h e^{-\frac{h}{\gamma}}-d^2 \gamma^2 e^{-\frac{h}{\gamma}}-d^2 \gamma h e^{-\frac{h}{\gamma}}}{d(d\gamma -1)^2} \\
    \frac{\gamma-\gamma e^{-\frac{h}{\gamma}} -h e^{-\frac{h}{\gamma}}}{\gamma} \\
    1-e^{-\frac{h}{\gamma}}
\end{bmatrix}$ \\
\hline
\end{tabular}
\end{table*}
\vspace{-0.1cm}
\subsection{Inter-sample constraints satisfaction}\label{MPC1 intersample Section }
\vspace{-0.3cm}
To achieve good tracking performance, especially in cases of long sampling periods, the MPC needs to account for the inter-sample behavior. The following lemma demonstrates that the MPC law in (\ref{MPC problem 1}) ensures the satisfaction of constraint (\ref{TV cons decoupled}) between sample times.
\begin{lem}
\label{intersample lemma}
    The MPC law resulting from the optimization problem (\ref{MPC problem 1}) guarantees satisfaction of the constraint (\ref{TV cons decoupled}) between intersample intervals. 
    \end{lem}
    \vspace{-0.45cm}
\begin{pf}
    The evolution of the acceleration of the system (\ref{decoupled outerloop dynamics}) in between sample times equals 
    \vspace{-0.1 cm}
\begin{multline}
\label{a(t)}
a_d(t)=\alpha(t-t_k)~a_d(t_k)+\beta(t-t_k)~\eta(t_k) \\
+[1-\alpha(t-t_k)-\beta(t-t_k)]~u(t_k),
\end{multline}
for $t \in [t_k,t_{k+1})$, where $\alpha(t)=e^{-\frac{t}{\gamma}}$, $\beta(t)=\frac{t}{\gamma}e^{-\frac{t}{\gamma}}$. Note that $0 < \alpha(t)\leq 1$, $0 < \beta(t)\leq e^{-1}$, and $0< \alpha(t)+\beta(t)\leq 1$. In accordance with \eqref{a(t)}, at each time instance $a_d(t)$ is a convex combination of $a_d(t_k),~\eta(t_k)$ and $u(t_k)$. Hence, from $|a_d(t_k)| \leq \Delta_{0|k}$, $|\eta(t_k)| \leq \Delta_{0|k}$ and $|u(t_k)| \leq \Delta_{0|k}$, we deduce that $|a_d(t)| \leq \Delta_{0|k}$ for all $t \in [t_k,t_{k+1})$. Because $\Delta_{0|k}\leq \Delta(t)$ for all $t \in [t_k,t_{k+1}]$ by definition, we can conclude that $|a_d(t)| \leq \Delta(t)$ for all $t$ within this interval.
\qed
\end{pf}

\vspace{-0.45cm}
\subsection{Converting the state constraints set into input constraints}\label{MPC1 constraint Section}
\vspace{-0.3cm}
The MPC problem (\ref{MPC problem 1}) includes two state constraints and one input constraint.
Terminal constraints are typically employed to guarantee stability and recursive feasibility in state-constrained linear receding horizon control schemes, but are often avoided due to their restrictive nature %limiting the controller's operational region
and complicating numerical optimization.
%in each MPC step. %(\cite{stateconstraintMPC}).
 In \cite{statetoinput}, a nonlinear mapping is derived that converts the state constraint set into an equivalent TV input constraint set for discrete-time linear systems. This approach anticipates the effect of current control actions on the future state and recalculates admissible control actions to prevent constraint violations. Using the subsequent two lemmas, the \textit{state} constraints (\ref{35b}) and (\ref{35c}) are transformed into equivalent \textit{input} constraints employing this method. The discrete-time dynamics of the states $a_d$ and $\eta$ of system \eqref{DT system dynamics} are formulated as follows:
\begin{align}
a_d(k+1)&=\alpha a_d(k)+\beta \eta(k)+[1-\alpha-\beta] u(k), \label{a(i+1)} \\
\eta(k+1)&=\alpha \eta(k)+[1-\alpha]u(k), \label{eta(i+1)} 
\end{align}
where $\alpha=e^{-\frac{h}{\gamma}}$,  $\beta=\frac{h}{\gamma}e^{-\frac{h}{\gamma}}$, satisfying $0 < \alpha\leq 1$, $0 < \beta\leq e^{-1}$, and $0< \alpha+\beta\leq 1$.
\begin{lem}
\label{ad TV lemma}
Consider the MPC problem defined in \eqref{MPC problem 1}. If the TV input constraint
\begin{equation}
    \Tilde{\Delta}^-_{i|k}\leq u_{i|k} \leq \Tilde{\Delta}^+_{i|k},\quad i\in \{0,1,..,N-1\}
\label{Deltatilde cons}
\end{equation}
holds, where 
\begin{align}
\Tilde{\Delta}^+_{i|k}&:=(1-\alpha-\beta)^{-1}[\Delta_{i+1|k}-(\alpha a_{d_{i|k}}+\beta \eta_{i|k})], \label{deltatilde +} \\
\Tilde{\Delta}^-_{i|k}&:=(1-\alpha-\beta)^{-1}[-\Delta_{i+1|k}-(\alpha a_{d_{i|k}}+\beta \eta_{i|k})], \label{deltatilde -}
\end{align}  
then it also holds that $|a_{d_{i+1|k}}| \leq \Delta_{i+1|k}$.
\end{lem}
\vspace{-0.5cm}
\begin{pf}
As mentioned, $(1-\alpha-\beta)$
  is non-negative. Multiplying $(1-\alpha-\beta)$ to \eqref{Deltatilde cons} and adding $(\alpha a_{d_{i|k}}+\beta \eta_{i|k})$ to all three terms of the resulting inequalities yields
\begin{equation*}
     -\Delta_{i+1|k} \leq \alpha a_{d_{i|k}}+\beta \eta_{i|k}+(1-\alpha-\beta)u_{i|k} \leq \Delta_{i+1|k}.
  \end{equation*} 
Using the dynamics of the state  $a_d$ in \eqref{a(i+1)}, it can be concluded that $|a_{d_{i+1|k}}| \leq \Delta_{i+1|k}$.
\qed
\end{pf}   

\begin{lem}
\label{eta TV lemma}
Consider the MPC problem defined in \eqref{MPC problem 1}. If the TV input constraint  
\begin{equation}
    \Bar{\Delta}^-_{i|k}\leq u_{i|k} \leq \Bar{\Delta}^+_{i|k},
\label{Deltabar cons}
\end{equation}
holds, where 
\begin{align}
\Bar{\Delta}^+_{i|k}&:=(1-\alpha)^{-1}[\Delta_{i+1|k}-\alpha \eta_{i|k}], \label{deltabar +} \\
\Bar{\Delta}^-_{i|k}&:=-(1-\alpha)^{-1}[\Delta_{i+1|k}+\alpha \eta_{i|k}], \label{deltabar -}
\end{align}    
then it implies that $|\eta_{i+1|k}| \leq \Delta_{i+1|k}.$
\end{lem}
\vspace{-0.5cm}
\begin{pf}
\iffalse
As mentioned, $1-\alpha$
  is a non-negative constant value. Multiplying $1-\alpha$ to \eqref{Deltabar cons} and adding $\alpha a_{d_{i|k}}$ to all three terms of the resulting inequalities yields
\begin{equation*}
     -\Delta_{i+1|k} \leq \alpha \eta_{i|k}+(1-\alpha)~ u_{i|k} \leq \Delta_{i+1|k}
  \end{equation*} 
Using the dynamics of the state  $\eta$ in \eqref{eta(i+1)},
\fi
Multiplying \( (1 - \alpha \)) to \eqref{Deltabar cons}, adding \( \alpha a_{d_{i|k}} \), and applying the same reasoning as in the proof of Lemma \ref{ad TV lemma}, yields $|\eta_{i+1|k}| \leq \Delta_{i+1|k}$.
\qed
\end{pf} 
\vspace{-0.5cm}
Leveraging Lemmas \ref{ad TV lemma} and \ref{eta TV lemma}, the two state constraints (\ref{35b}) and (\ref{35c}) are replaced by their corresponding input constraints. As a result, all three constraints (\ref{35a})-(\ref{35c}) can be unified into a single input constraint as
\begin{equation}
\label{input constraint}
u_{i|k} \in [u_{{min}_{i|k}},u_{{max}_{i|k}}],~~~i\in \{0,1,..,N-1\},
\end{equation}
where
\begin{align}
u_{{min}_{i|k}}&= \max(-\Delta_{i|k}, \Tilde{\Delta}^-_{i|k}, \Bar{\Delta}^-_{i|k} ), \label{umin} \\
u_{{max}_{i|k}}&= \min(+\Delta_{i|k}, \Tilde{\Delta}^+_{i|k}, \Bar{\Delta}^+ _{i|k}), \label{umax} 
\end{align}     
and $\Tilde{\Delta}^+_{i|k}$, $\Tilde{\Delta}^-_{i|k}$, $\Bar{\Delta}^+_{i|k}$ and $\Bar{\Delta}^-_{i|k}$ defined in \eqref{deltatilde +}, \eqref{deltatilde -}, \eqref{deltabar +} and \eqref{deltabar -}.
The necessary condition for the feasibility of the MPC problem with input constraint (\ref{input constraint}) is established in Theorem \ref{TV theorem}, ensuring that the input set is non-empty.
\begin{thm}
\label{TV theorem}
If $h$ and $\gamma$ are appropriately chosen such that the condition
\begin{equation}
    \Delta (k+1) > e^{-\frac{h}{\gamma}} \left(1 + \frac{h}{\gamma}\right) \Delta(k), \quad \forall k \in \mathbb{N}_0,  
    \label{h/gam cons}
\end{equation}
holds for a given finite-time trajectory with \(\Delta(k) = \Delta_{0|k}\) in (\ref{Deltak definition}), then the unified input constraint in (\ref{input constraint}) is non-empty, and the MPC controller \eqref{MPC problem 1} with the unified input constraint (\ref{input constraint}) will be feasible for all $x \in \mathbb{R}^4$.
\end{thm}
 \vspace{-0.6cm}
\begin{pf}
See Appendix \ref{TV theorem proof}.
\qed
\end{pf}

\vspace{-0.4cm}
\subsection{Stability Guarantee} \label{MPC1 stabilityt Section}
\vspace{-0.3cm}
This study aims to guarantee UGAS of the tracking error system for the proposed MPC strategy. In \cite{Alexpaper}, a stabilizing input constraint is added to ensure the existence of a Lyapunov function, thus guaranteeing UGAS of the outer loop. However, adding a stabilizing input constraint compromises performance through the
introduction of conservatism.
 Given that the system in (\ref{decoupled outerloop dynamics}) is marginally stable, an appropriate cost function can be used to find a Lyapunov function that guarantees the UGAS of the closed-loop system. As outlined in Section \ref{section 2 preliminary}, a finite-horizon MPC is proposed in \cite{neutrallystable}, which globally stabilizes marginally stable linear systems subject to input constraints by using a non-quadratic function as the terminal cost, consisting both cubic and quadratic functions of the state. This method is tailored to quadcopter setting here to ensure UGAS of the outer-loop problem without compromising performance. 
  Since the MPC problem (\ref{MPC problem 1}) deals with TV input constraints, it is necessary to derive conditions that ensure the validity of the proof of Theorem \ref{GAS theorem} for TV input constraints. Thus, a smaller time-invariant input set is derived in the following lemma, which is only used for the purpose of the stability proof. This set serves as the saturation function (\ref{sat function}), allowing Theorem \ref{Thoery small gain} %and \ref{GAS theorem} 
  to be applied to determine the Lyapunov function.
  %and terminal cost, respectively.
\begin{lem}
\label{smallest input set lemma}
Consider the TV input set (\ref{input constraint}) 
\begin{equation*}
    \mathscr{U}_{i|k}=\{ u \in \mathbb{R}|~ 
u_{{min}_{i|k}} \leq u \leq u_{{max}_{i|k}}\},
\end{equation*}
with $u_{{min}_{i|k}}$ and $u_{{max}_{i|k}}$ defined in (\ref{umin}) and (\ref{umax}), then the following time-invariant input set  is a subset of $ \mathscr{U}_{i|k}$ at all sample times $k\in \mathbb{N}_0$:
\begin{equation}
\label{Time_invarient input set}
    \mathscr{U}=\{ u \in \mathbb{R}|~ 
-\Delta^* \leq u \leq \Delta^*\},
\end{equation}
where $\Delta^*$ is a positive constant value obtained by
\begin{equation}
\label{delta*}
    \Delta^*= \min(\Tilde{\Delta}^+_{min} ,~ \bar{\Delta}^+_{min},~\Delta_{min})~,
\end{equation}    
and
\begin{align}
        \Tilde{\Delta}^+_{min}&=(1-\alpha-\beta)^{-1} \min_{k}[\Delta(k+1)-(\alpha+\beta)\Delta(k)], \nonumber\\
        \bar{\Delta}^+_{min}&=(1-\alpha)^{-1}\min_{k} (\Delta(k+1)-\alpha \Delta(k)),\nonumber\\
        \Delta_{min}&=\min_{k} \Delta(k),
   \label{delta star bounds}     
   \end{align}
with $\Delta(k)=\Delta_{0|k}$ in (\ref{Deltak definition}).
\end{lem}
\vspace{-0.4cm}
\begin{pf}
    See Appendix \ref{lemma 5 proof}
    \qed
\end{pf}
\vspace{-0.5cm}
Theorem \ref{MPC1 stablity theorem} below presents the main contribution of this section, as it describes the design of an MPC problem with TV input constraints, ensuring UGAS and inter-sample constraint guarantees.

\begin{thm}
\label{MPC1 stablity theorem}
    Consider the closed-loop system \eqref{DT system dynamics}, where the open-loop system is marginally stable and the pair $(A_d,B_d)$ is controllable, with the control law $k_N(x_k)$ as in (\ref{kn}), resulting from the optimization problem
\begin{align}   
\label{stable MPC problem 1_final}
    \min_{U_k}~&~ J(x_k,U_k) = V(x_{N|k})+\sum_{i=0}^{N-1}(x_{i|k}^{\T}Qx_{i|k}+u_{i|k}^{\T}R ~u_{i|k})\nonumber\\
    \mathrm{s.t.}~&~ x_{0|k}=~x_k,\\
    &~x_{i+1|k}=~A_d x_{i|k}+B_d u_{i|k},~i\in \{0,1,..,N-1\},\nonumber\\
    &~u_{{min}_{i|k}} \leq u_{i|k} \leq u_{{max}_{i|k}},~~~i\in \{0,1,..,N-1\},\nonumber
\end{align} 
where $u_{{min}_{i|k}}$ and $u_{{max}_{i|k}}$ defined in (\ref{umin}) and (\ref{umax}) and
\begin{equation*}
\hspace*{0cm} % Adjust the value as needed
\begin{aligned}
    & V(x)=\Theta W(x)=\Theta\Bigr[x^{\T} M_q x+\lambda(x^{\T} M_c x)^{3/2}\Bigr],
\end{aligned}
\end{equation*}
with $M_c$ and $M_q$ are as in \eqref{Mc} and \eqref{Mq}, $N \in \mathbb{N}$ is the prediction horizon, $Q$ and $R$ are positive definite matrices, and $\Theta$ is as specified in \eqref{theta}.
Furthermore, let $h$ and $\gamma$ be chosen appropriately such that
\begin{equation*}
\Delta (k+1) > e^{-\frac{h}{\gamma}}\left(1+\frac{h}{\gamma}\right)\Delta (k),\quad\forall k \in \mathbb{N}_0,
    \end{equation*}
    for the given trajectory, with $\Delta(k)=\Delta_{0|k}$ in (\ref{Deltak definition}). Then for any positive integer $N$, the closed-loop system $x^+=A_dx+B_dk_N(x)$ is uniformly globally asymptotically stable for 
\begin{equation*}
        \lambda=\frac{2 \kappa L_u \sigma_{max}(A_d^{\T} M_q B_d)}{\sqrt{\lambda_{min}(M_c)}},
    \end{equation*}
    where $\kappa$ satisfies (\ref{kappa}),  and $L_u$ is chosen such that  $\Delta^* L_u > 1$ with $\Delta^*$ defined in (\ref{delta*}) and (\ref{delta star bounds}).
    \end{thm}
    \vspace{-0.5cm}
    \begin{pf}
 Consider the saturation function (\ref{sat function}) with $u_{\mathrm{max}}=\Delta^*$.  
According to Theorem \ref{Thoery small gain}, the local control law $ k_f:\mathbb{R}^{4}  \rightarrow \mathbb{R}$, defined as $k_f(x)=\mathrm{sat}(Kx)$ with
the small gain control $K$ in (\ref{smallgain}), ensures UGAS of the closed-loop system, 
with the global Lyapunov function defined in \eqref{LyapFunc}. 
Knowing that $k_f(x)=\mathrm{sat}(Kx)$, we can infer that 
$k_f(x) \in \mathscr{U}$ with $\mathscr{U}$ defined in (\ref{Time_invarient input set}).
By applying Lemma \ref{smallest input set lemma}, it can be inferred that $\mathscr{U} \subset \mathscr{U}_{i|k}$. Demonstrating that $k_f \in \mathscr{U}_{i|k}$, the assumptions of Lemma \ref{lemma 1} and Theorem \ref{GAS theorem} remains valid for a saturation function with  $u_{\mathrm{max}}=\Delta^*$. Therefore, the MPC problem (\ref{stable MPC problem 1_final}) demonstrates UGAS if $L_u$ is chosen such that  $\Delta^* L_u > 1$.
\qed
\end{pf}

\vspace{-0.6cm}
\section{Stability of the Continuous-time System} \label{section Stability of the continuous system}
\vspace{-0.3cm}
In Sections \ref{MPC1 Section}, UGAS is demonstrated for the proposed MPC strategy. The states in the discrete-time dynamics \eqref{DT system dynamics} exactly correspond to the states of the continuous-time dynamics (\ref{decoupled outerloop dynamics}) due to the exact discretization used. However, it is essential to take into account the behavior of the continuous-time system between sampling times to conclude UGAS for the continuous-time system.
The continuous-time linear system (\ref{Main outerloop dynamics}) can be written as 
\begin{equation}
    \Dot{x}=Ax(t)+Bu(t)
    \label{CT system coupled}
\end{equation}
where $x(t)=[\Tilde{p}(t) ~ \Tilde{v}(t)~ a_{d}(t)~ \eta(t)]^{\T} \in \mathbb{R}^{12} $ and the input 
\begin{equation}
\label{CT input}
    u(t)=
    \begin{bmatrix}
        u_{x_{MPC}}(x(kh))\\
        u_{y_{MPC}}(x(kh))\\
        u_{z_{MPC}}(x(kh))
    \end{bmatrix}, ~~ \forall t\in[kh,kh+h),
\end{equation}
is generated by three separate MPC problems. Then for all $t\in[kh,kh+h]$ :
\begin{equation*}
    x(t)=e^{A(t-kh)} x(kh)+   \int_{kh}^{t} e^{A(t-\tau)} \,d\tau B~u(kh).
\end{equation*}
Knowing that matrix $A$ represents the state matrix of a marginally stable system, we can infer that
\begin{align*}
    \|x(t)\|=&\|e^{A(t-kh)} x(kh)+   \int_{kh}^{t} e^{A(t-\tau)} \,d\tau B~u(kh)\|  \\ 
     \leq&\|e^{A(t-kh)}\| \|x(kh)\|+   \|\int_{kh}^{t} e^{A(t-\tau)} \,d\tau B\|\|u(kh)\|\\
    \leq& c_1  \|x(kh)\| + c_2 \|u(kh)\|.
\end{align*}
In the formulated MPC strategy, $u(kh)$ is constrained such that $\|u(kh)\| \leq \Delta (t)=\frac{1}{\sqrt{3}}\rho (t)$ for all $k\in \mathbb{N}_0$, where 
\begin{equation*}
    \rho (t)=\min(\bar{T}(t)-\delta~,~T_{max}-\bar{T}(t)) \leq \max\limits_{t}\bar{T}(t) \leq T_{\mathrm{max}}.
\end{equation*}
This bound on input, together with the UGAS property of $x(kh)$
 in discrete-time, ensures the states of the continuous-time system remain bounded between sampling times. Hence, the closed-loop system \eqref{CT system coupled} with ZOH-input generated by MPC controllers, is guaranteed to be UGAS.
\vspace{-0.2cm}
\section{Cascade Trajectory Tracking Controller} \label{cascade section}
\vspace{-0.2cm}
The MPC controller in the outer-loop problem generates a desired acceleration, which has been demonstrated to be asymptotically stable. The desired acceleration is converted into a desired attitude, which is tracked by the attitude controller in the inner loop. The inner-loop controller has been proven to be ULES and UaGAS in \cite{ErijenNLcontroller}. To conclude the stability of the closed-loop system, it is necessary to demonstrate stability for the full cascade system. The closed-loop system resulting from the dynamics (\ref{dynamics}) and reference (\ref{reference}), with inputs defined in (\ref{thrust}), (\ref{tau}), and (\ref{CT input}), is expressed as:
\begin{subequations} \label{cascade dynamics}
\begin{align}
\label{position error cascade}
  {\Dot{\Tilde{p}}}&=\Tilde{v},\\
\label{velocity error cascade}
   \Dot{\Tilde{v}}&=-D\Tilde{v}+a_d+TR(I-R_e^{\T})~e_3,\\
\label{ad dynamic cascade}
   \Dot{a_d}&=-\frac{1}{\gamma}(a_d+\eta),\\
\label{eta dynamic cascade}
   \Dot{\eta}&=-\frac{1}{\gamma}(\eta+u),\\
\label{Re dynamic cascade}
   \Dot{R_e}&=~R_e S(\omega_e),\\
\label{omega dynamic cascade}
   J\Dot{\omega_e}&=-K_\omega \omega_e + K_R \sum_{i=1}^{3} k_i (e_i \times R_e^{\T} e_i),
\end{align}
\end{subequations}
which is a cascade of the systems (\ref{position error cascade})-(\ref{eta dynamic cascade}) and (\ref{Re dynamic cascade})-(\ref{omega dynamic cascade}).
\begin{thm} 
    \label{cascade theorem}
The origin $(\Tilde{p}, \Tilde{v}, R_e, \omega_e)=(0,0,I,0)$ of (\ref{cascade dynamics}) is UaGAS \cite{Alexpaper}.
\end{thm}
\vspace{-0.5cm}
\begin{pf}
Given that the outer-loop dynamic \eqref{CT system coupled} is UGAS on $\mathbb{R}^{12}$ and the inner-loop dynamics is UaGAS on $SO(3) \times \mathbb{R}^3$, \cite{Alexpaper} demonstrates that if the solution remains bounded, UaGAS for (\ref{cascade dynamics}) is guaranteed. The dynamics (\ref{Re dynamic cascade})-(\ref{omega dynamic cascade}) are bounded since the inner-loop dynamics in (\ref{cascade dynamics}) have been proven to be UaGAS on $SO(3) \times \mathbb{R}^3$ in \cite{ErijenNLcontroller}.
Therefore, demonstrating the boundedness of solutions for (\ref{position error cascade})-(\ref{eta dynamic cascade}) suffices to ensure the stability of the full cascade system. For further details, refer to \cite{Alexpaper}. The boundedness of solutions for (\ref{position error cascade})-(\ref{eta dynamic cascade}) is proved in \mbox{Appendix \ref{app1}}. \qed
\end{pf}
\vspace{-0.5cm}
Finally, it is demonstrated that achieving UaGAS for (\ref{cascade dynamics}) provides a solution to Problem \ref{problem1}. Having UaGAS, as stated in Theorem \ref{cascade theorem}, guarantees robustness against uniformly bounded perturbations, as described in \cite{Alexpaper}.
\vspace{-0.05cm}
\begin{corollary}
The controller, which uses the inputs derived from (\ref{thrust}), (\ref{tau}), and (\ref{CT input}), solves Problem 1.
\end{corollary}
\vspace{-0.45cm}
\begin{pf}
 See \cite{Alexpaper}.
\end{pf}

\vspace{-0.55cm}
\section{Numerical and High-Fidelity Case Studies} \label{case studies section}
\vspace{-0.3cm}
This section includes numerical case studies, using (\ref{dynamics}) as the quadcopter model, as well as high-fidelity case studies to verify the effectiveness of the proposed cascade control schemes. In particular, the dynamic in (\ref{DT system dynamics}) is implemented for the state prediction process to obtain the solution of the %optimal control problem (OCP)
MPC problem
 in the outer loop.
\vspace{-0.1 cm}
\subsection{Numerical case study} \label{Numerical example section}
 \vspace{-0.3cm}
In this section, we compare the proposed cascade control strategy with the controller from \cite{Alexpaper} through a numerical example.  The simulations are conducted in
MATLAB  for a trajectory of $25$ seconds. \enquote{\textit{Proposed MPC}} refers to the MPC strategy formulated in Section \ref{MPC1 Section}, while the MPC strategy from \cite{Alexpaper} is labeled as \enquote{\textit{Baseline MPC}} in the resulting plots. The nonlinear controller from Section \ref{section NLC} is employed for inner-loop control in both cases.
Following \cite{Alexpaper}, the dynamics in (\ref{dynamics}) for the simulation are based on the parameters used in \cite{Romero2022}.
%of an in-house designed race drone by the Robotics and Perception Group at the University of Zurich used in \cite{Romero2022}.
These parameters are  $g = \unit[9.81]{m/s^2}$, $J = \unit[\mathrm{diag}(2.5, 2.1, 4.3)]{gm^2}$, $D =\unit[\mathrm{diag}(0.26, 0.28, 0.42)]{kg/s}$, $\tau_g = [0, 0, 0]^\top$, $A = 0.1I$, $C = 0.5I$, and $ T_{\mathrm{max}} = 45.21~\mathrm{m/s}^2$.

The challenging reference trajectory is considered as 
\begin{align}
        \Bar{p}(t)&=[2\cos(4t)~~~2\sin(4t)~~~-10+2\sin(2t)]^{\T}, \notag\\
\Bar{\psi}(t)&=0.2t,
\label{simulationtrajectory}
\end{align}
where $\Bar{p}(t)$ specifies the position reference and $\Bar{\psi}(t)$ represents the heading reference, indicating the angle between the projection of $x_B$ onto the $x_G - y_G$ plane and the $x_G$ axis. This reference trajectory completely determines the states and inputs of the reference model in (\ref{reference}), based on the differential flatness property of the quadcopter \cite{Faessler2017}.

The simulation starts from the same initial conditions as in the numerical example from \cite{Alexpaper}.
\iffalse
The simulation starts from the following initial conditions:
\begin{align*}
    p(0)&= [1.5\Bar{p}_x(0)~~0.75\Bar{p}_y(0)~~\Bar{p}_z(0)]^{\T},~~~~~~
    v(0)=\bar{v}(0),\\
    R(0)&= R_x\bigg(\frac{170\pi}{180}\bigg)R_y\bigg(\frac{30\pi}{180}\bigg)R_z\bigg(\frac{20\pi}{180}\bigg),~
    \omega(0)=\bar{\omega}(0),
\end{align*}
where $R_x$ , $R_y$ and $R_z$ represents the rotation matrix around the x,  $y$ and $z$ axes, respectively.
\fi
The inner-loop gains are $K_{\omega} = 30J$, $K_{R} = 70J$, and $k = [4.5,~5,~5.5]$ in the simulation. The outer-loop parameters are $h=\unit[0.05]{s}$, $\gamma = 0.1$, $N = 20$, $Q = \mathrm{diag}(100, 1, 1, 1)$, and $R = 0.01$.
\iffalse
The simulations are based on the controller parameters  listed in Table \ref{tab:control param table}.
\begin{table}[!h]
\caption{Control parameters \label{tab:control param table}}
\centering
\begin{tabular}{|l|c|c|}
\hline
\textbf{Description} & \textbf{Symbol} & \textbf{Value}\\
\hline
\textbf{Inner-loop} & & \\
\hline
Feedback gain angular velocity & $K_{\omega}$ & $30J$\\
\hline
Feedback gain attitude & $K_{R}$ & $70J$\\
\hline
Feedback gain attitude & $k$ & $[4.5,~5,~5.5]$\\
\hline
\hline
\textbf{Outer-loop} & & \\
\hline
MPC sample time & $h$ & $0.05$ [s] \\
\hline
Model parameter & $\gamma$ & $0.1$ \\
\hline
Control horizon & $N$ & $20$\\
\hline
State cost matrix & $Q$ & diag$(100,1,1,1)$\\
\hline
Inpute cost matrix & $R$ & $0.01$\\
\hline
\end{tabular}
\end{table}

\begin{table}[!h]
\caption{Quadcopter parameters \label{tab:param table}}
\centering
\begin{tabular}{|c|c| c|}
\hline
\textbf{Description} & \textbf{Sym} & \textbf{Value}\\
\hline
\hline
Gravitational acceleration & $g$ & $9.81~ [\mathrm{m}/\mathrm{s}^2]$\\
\hline
Inertia matrix & $J$ & diag$(2.5, 2.1, 4.3)$ $[\mathrm{gm}^2]$\\
\hline
Translational drag coeffs & $D$ & diag$(0.26,0.28,0.42)$ [kg/s]  \\
\hline
Gyroscopic torques & $\tau_g$ & $[0, 0, 0]^{\T}$\\
\hline
Cross drag coefficients & $A$ & $0.1I$\\
\hline
Rotational drag coeffs& $C$ & $0.5I$\\
\hline
Max. thrust (mass-norm.) & $T_{\mathrm{max}}$ &  $45.21~[\mathrm{m}/\mathrm{s}^2]$\\
\hline
\end{tabular}
\end{table}
\fi
 To speed up the evaluation, CasADi \cite{Andersson:2019} is used for solving the MPC problems, chosen for its efficiency in nonlinear optimization and compatibility with C++, Python, and MATLAB. The multiple shooting method is used in the MPC function code to improve convergence. %(\cite{casadi_docs}).
 
\begin{figure}[!t]
\centering
\includegraphics[width=2.2in]{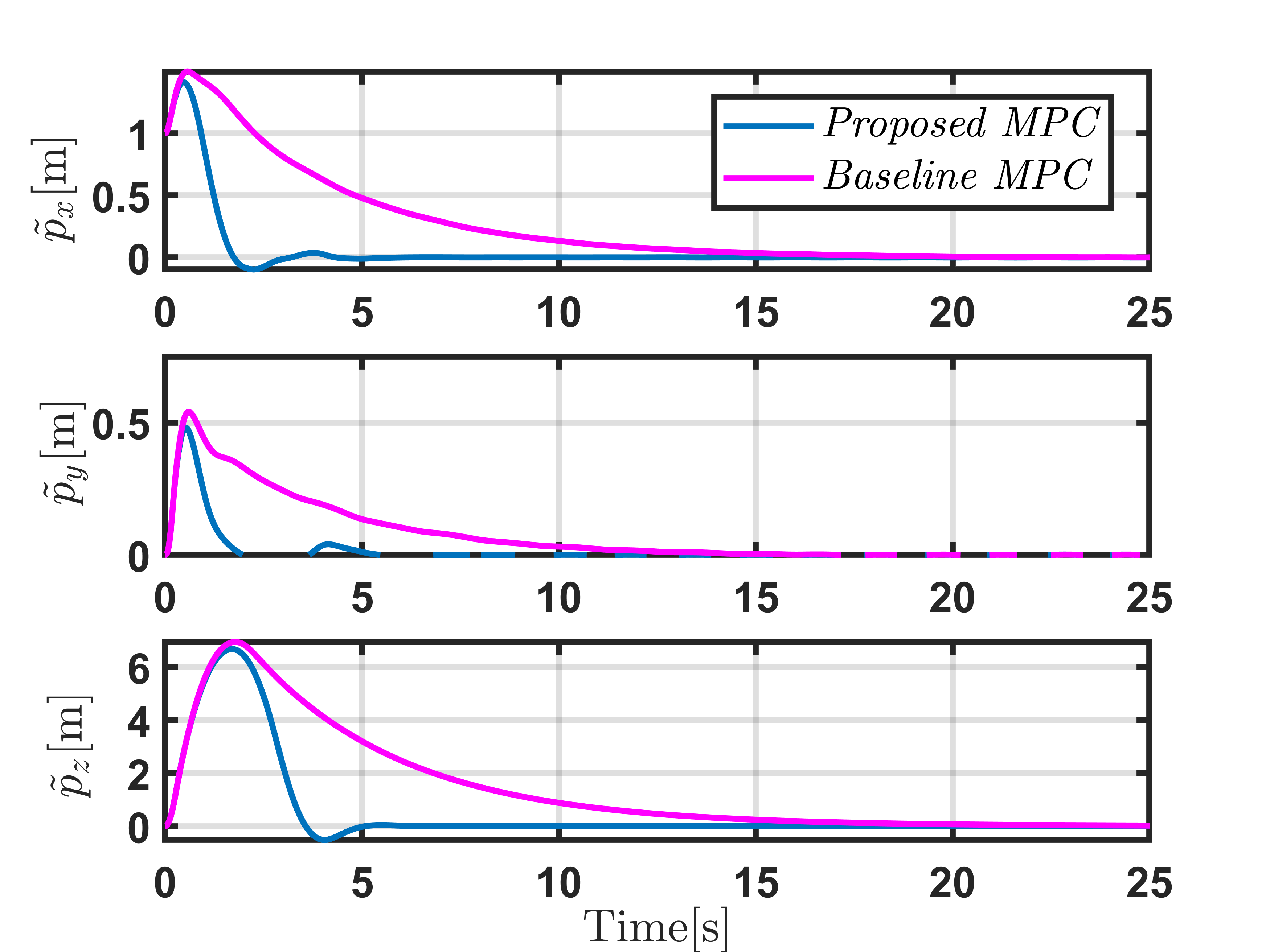}
\caption{Position errors $\Tilde{p}=[\Tilde{p}_x, \Tilde{p}_y, \Tilde{p}_z]$ for each axes.}
\label{error sim}
\end{figure}

\begin{figure}
\centering
\subfloat[]{\includegraphics[width=1.6in]{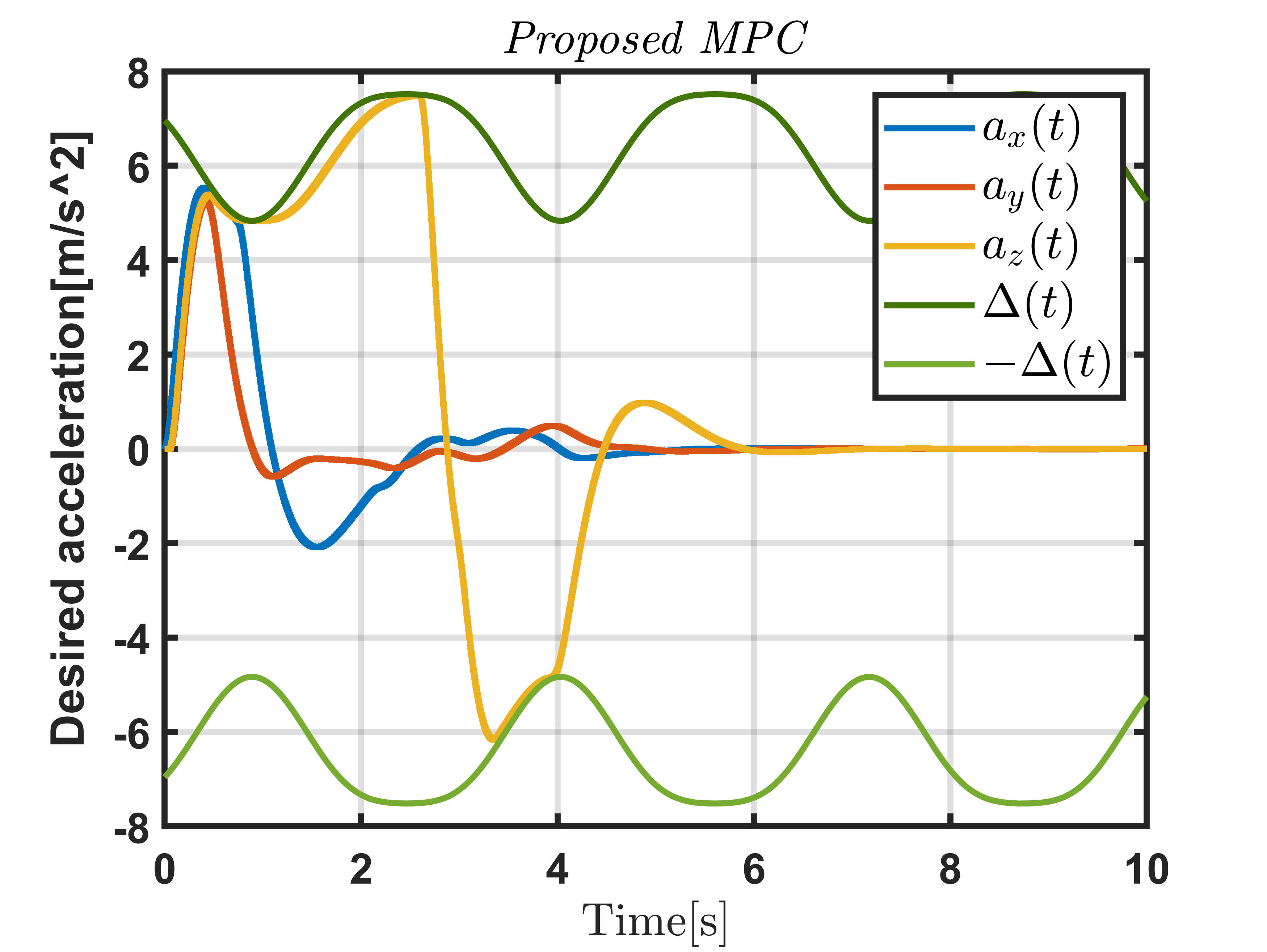}%
\label{MPC 1 ad}}
\hfil
\subfloat[]{\includegraphics[width=1.6in]{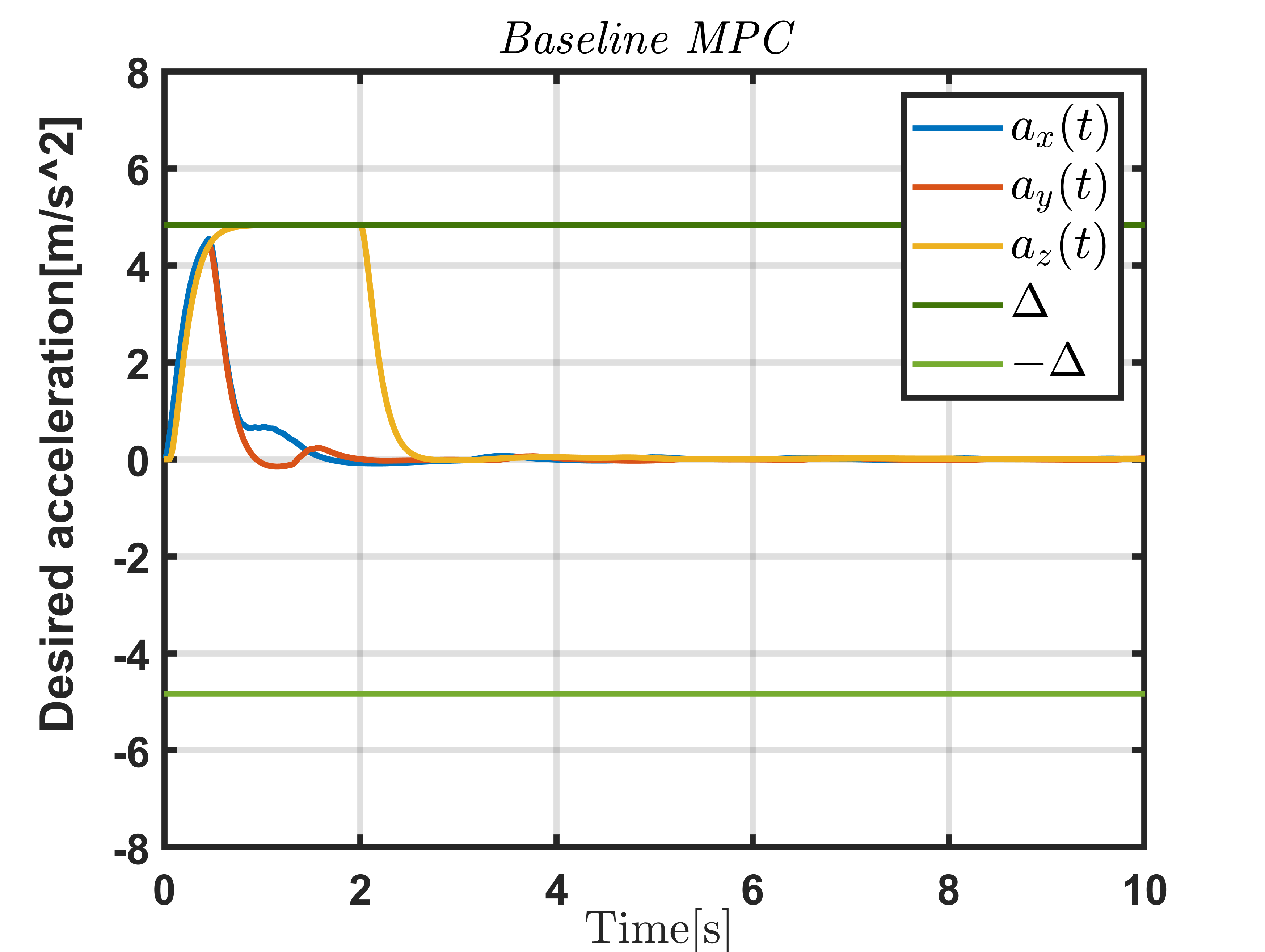}%
\label{MPC 3 ad}}
\caption{Desired accelerations in $x$, $y$, and $z$ during the first 10 seconds of the simulations. (a) shows that the \textit{Proposed MPC} keeps $a_{d,i}$ within $- \Delta(t)$ to $\Delta(t)$, satisfying constraint (\ref{TV cons decoupled}). (b) shows the \textit{Baseline MPC}, where $a_{d,i}$ are constrained within $- \Delta$ to $\Delta$, reflecting its conservative nature.}
\label{ad_sim}
\end{figure}

Fig. \ref{error sim} shows that \textit{Proposed MPC} outperforms \textit{Baseline MPC}, highlighting its effectiveness in reducing conservatism and improving performance. 
The Root Mean Square Error (RMSE) for the \textit{Proposed MPC} and \textit{Baseline MPC} are \unit[[0.26, 0.07, 1.75]]{m} and \unit[[0.48, 0.14, 2.43]]{m}, respectively. The \textit{Proposed MPC} executes in \unit[3.3]{ms} on average, compared to  \unit[0.25]{ms} for the \textit{Baseline MPC}. This increase in execution time is due to the fact that the \textit{Proposed MPC} is not a QP problem. However, the execution time in both cases is significantly below the sampling period of $h = \unit[50]{ms}$.
 These results show that the \textit{Proposed MPC} achieves a significantly lower RMSE compared to the \textit{Baseline MPC}, but at the cost of increased computational time.
Fig. \ref{ad_sim} illustrates the desired acceleration in the $x$, $y$, and $z$ directions for both cases. It is evident that \textit{Baseline MPC} applies more conservative constraints on $a_d$, showing that relaxing this constraint is instrumental in obtaining high performance.  Both schemes satisfy  UGAS property.
\subsection{High-Fidelity Case Study}
\label{HF section}
 \vspace{-0.3cm}
In this section we present numerical simulations in the Avular high-fidelity simulation environment for a quadcopter within MATLAB, aimed at validating the proposed cascade control strategy. The results are compared with a reference cascade control system that uses multiple PID controllers for altitude, horizontal position, and attitude control. Avular's simulation environment incorporates the dynamics specified in (\ref{dynamics}) and also includes motor dynamics, sensor measurements, and state estimation to closely mimic real-world conditions. The addition of artificial noise in sensor measurements and the subsequent state estimation introduces extra uncertainty. 
\enquote{\textit{Proposed cascade}} refers to the control strategies formulated in Sections \ref{section NLC} and \ref{MPC1 Section}, whereas the cascade PID-based controller is labeled as \enquote{\textit{PID}} in the resulting plots.

\begin{figure}
\centering
\includegraphics[width=2.2in]{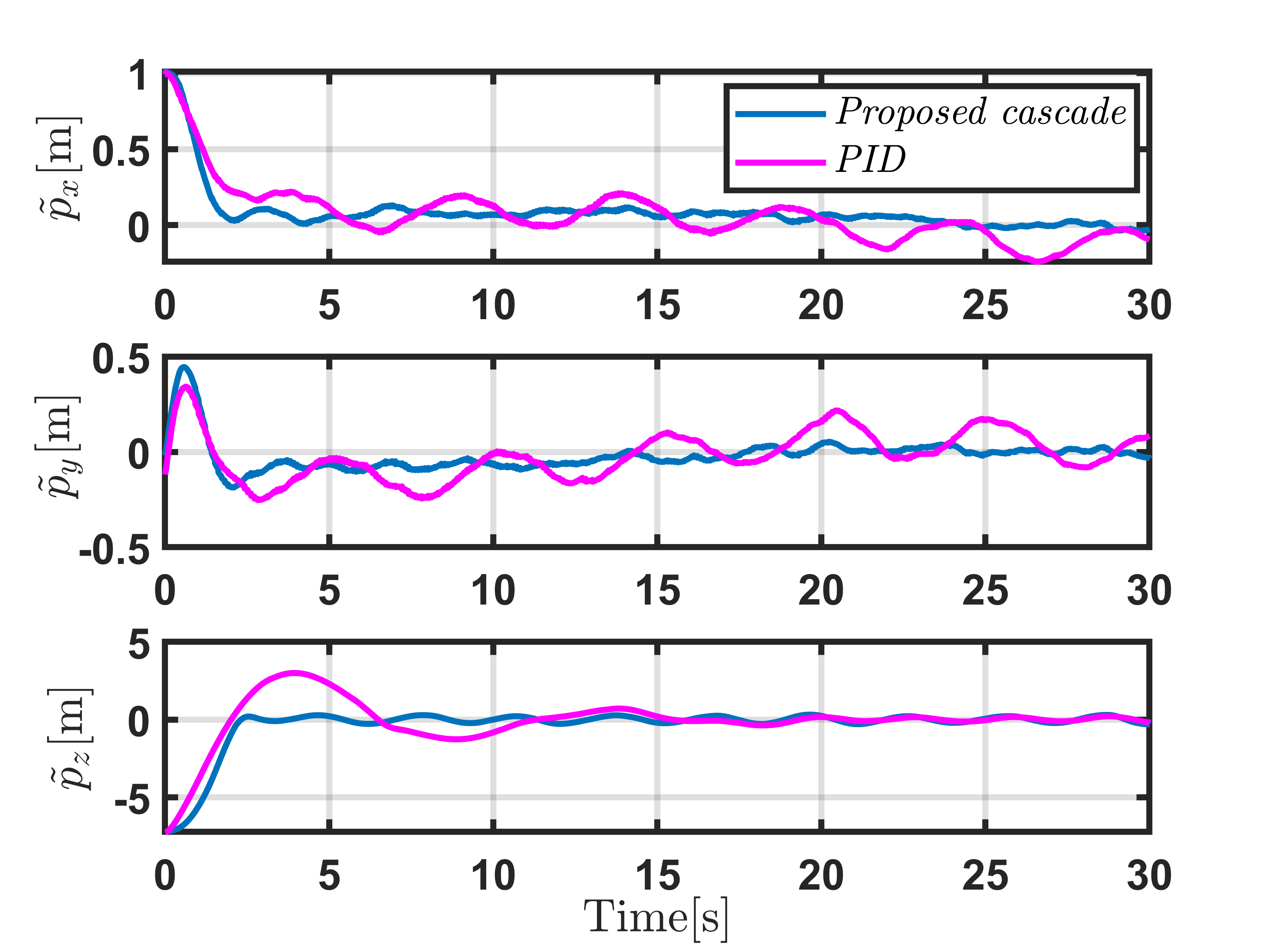}
\caption{Position errors $\Tilde{p}=[\Tilde{p}_x, \Tilde{p}_y, \Tilde{p}_z]$ for each axes, using the Avular high-fedility simulation environment.}
\label{error sim 2}
\end{figure}

The Avular Vertex One parameters, utilized in the simulations are $g = \unit[9.81]{m/s^2}$, $J =\unit[\mathrm{diag}(0.08, 0.08, 0.3)]{kgm^2}$, $D =\unit[\mathrm{diag}(0.15,0.15,0.15)]{kg/s}$, $\tau_g = [0, 0, 0]^\top$, $A = 0.1I$, $C = 0.5I$, $T_{\mathrm{max}} = \unit[80]{N}$ and, $m=\unit[2.8]{kg}$.
The inner-loop parameters are set to the same values as those in Section \ref{Numerical example section}. The outer-loop parameters are $h=\unit[0.1]{s}$, $\gamma = 0.1$, $N = 20$, $Q = \mathrm{diag}(100, 1, 1, 1)$, and $R =1$.
The simulations are conducted for 30 seconds and the reference trajectory is considered as
\begin{align}
        \Bar{p}(t)&=[\cos(0.4\pi t)~~\sin(0.4\pi t)~~-8+\sin(0.2\pi t)]^{\T}, \notag\\
\Bar{\psi}(t)&=0.2t,
\label{simulationtrajectory 2}
\end{align}
\begin{figure}[!t]
\centering
\subfloat[]{\includegraphics[width=1.6in]{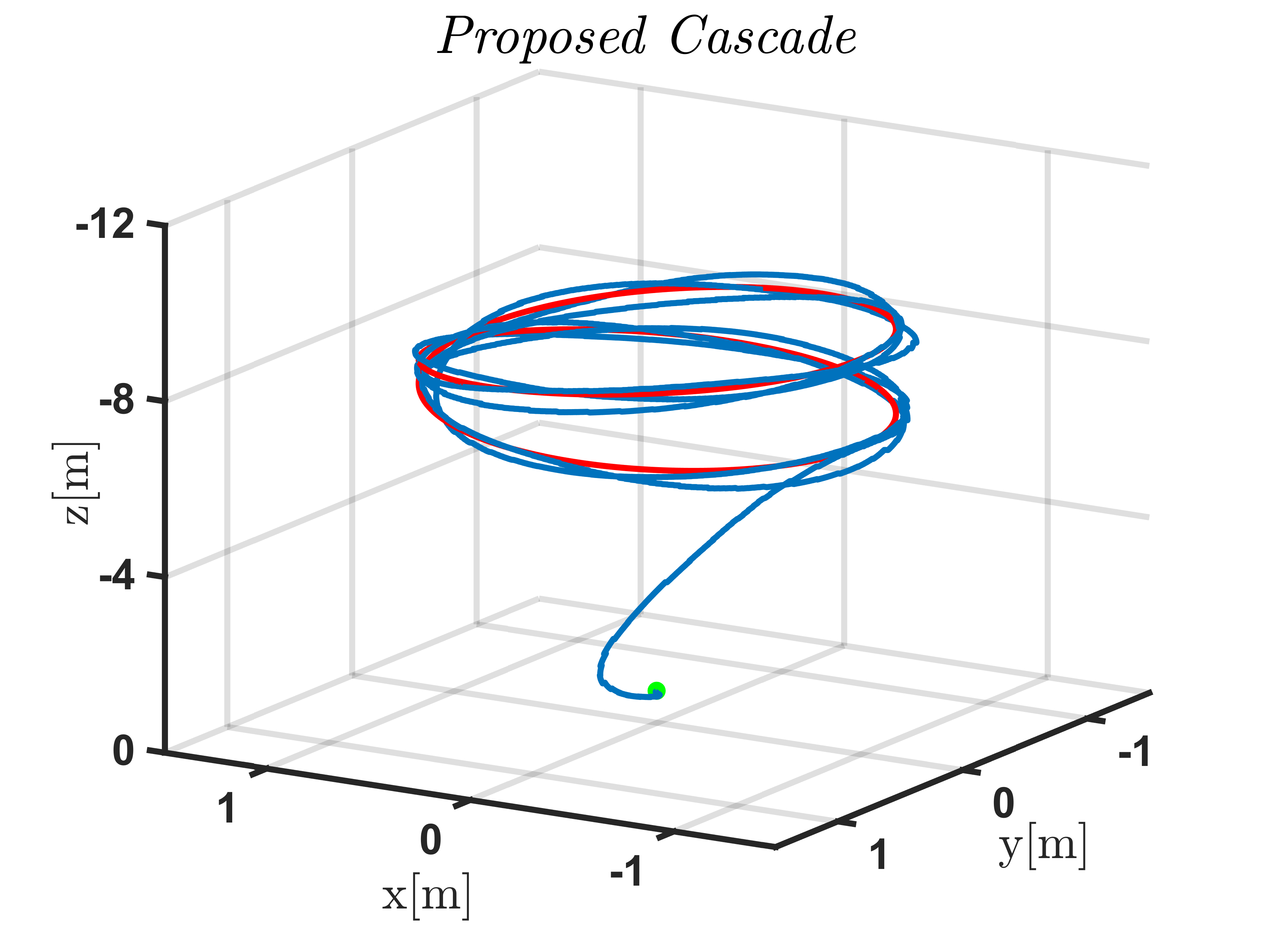}%
\label{MPC 1 trajetory sim2}}
\hfil
\subfloat[]{\includegraphics[width=1.6in]{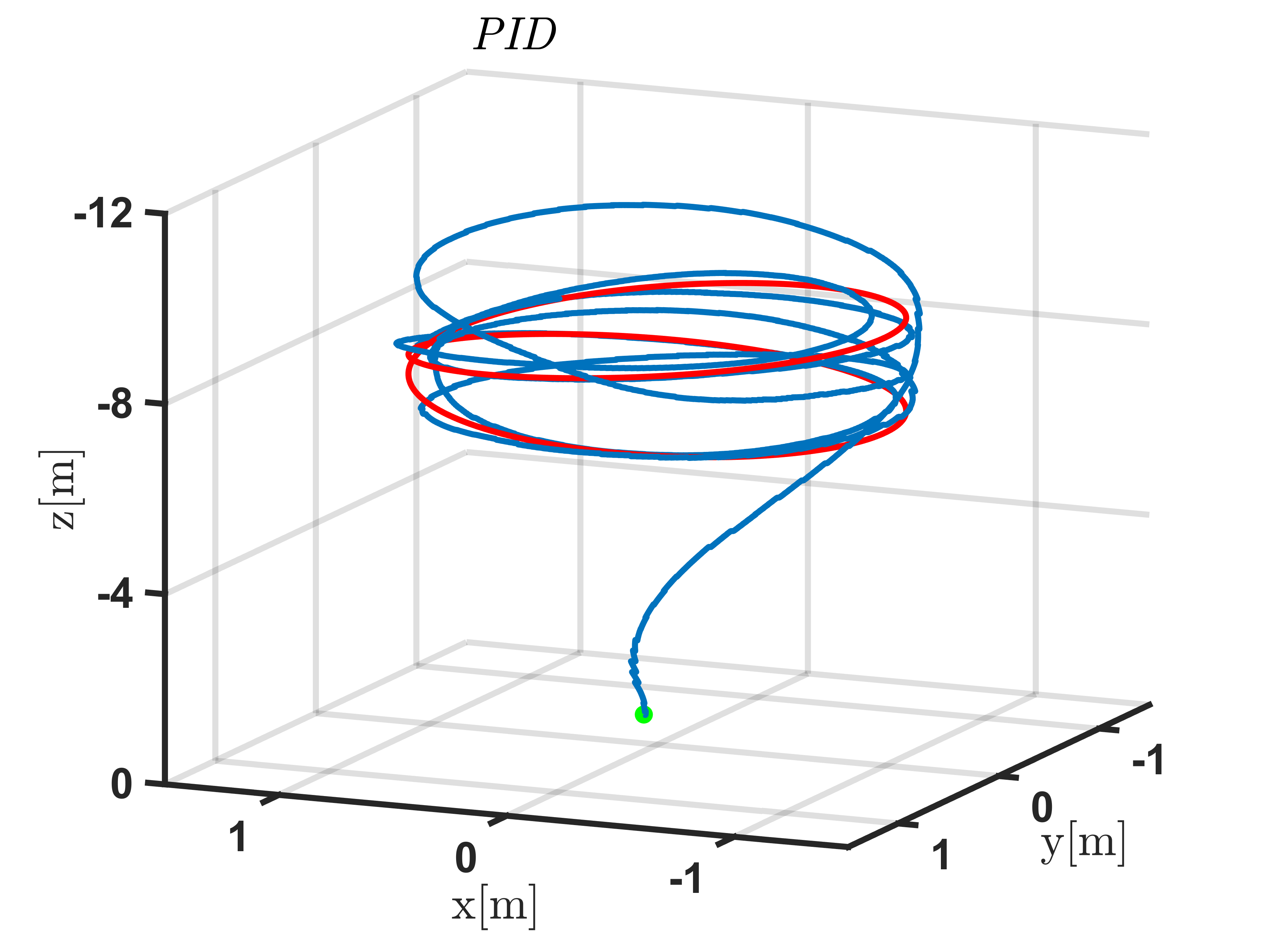}%
\label{PID trajetory sim2}}
\caption{3D plot of the reference and actual trajectories, using the Avular high-fedility simulation environment. The reference trajectory is indicated in red across both cases.}
\label{trajectories_sim2}
\end{figure}
Fig. \ref{trajectories_sim2} illustrates the reference and resulting position trajectories. As demonstrated by the trajectory plots and further validated by the position errors in Fig \ref{error sim 2}, \textit{Proposed cascade} outperforms \textit{PID}, highlighting its ability to effectively reduce overshoot in the $z$-direction step response, thanks to the predictive capability of MPC in anticipating future system behavior.

\begin{figure}
\centering
\includegraphics[width=2.2in]{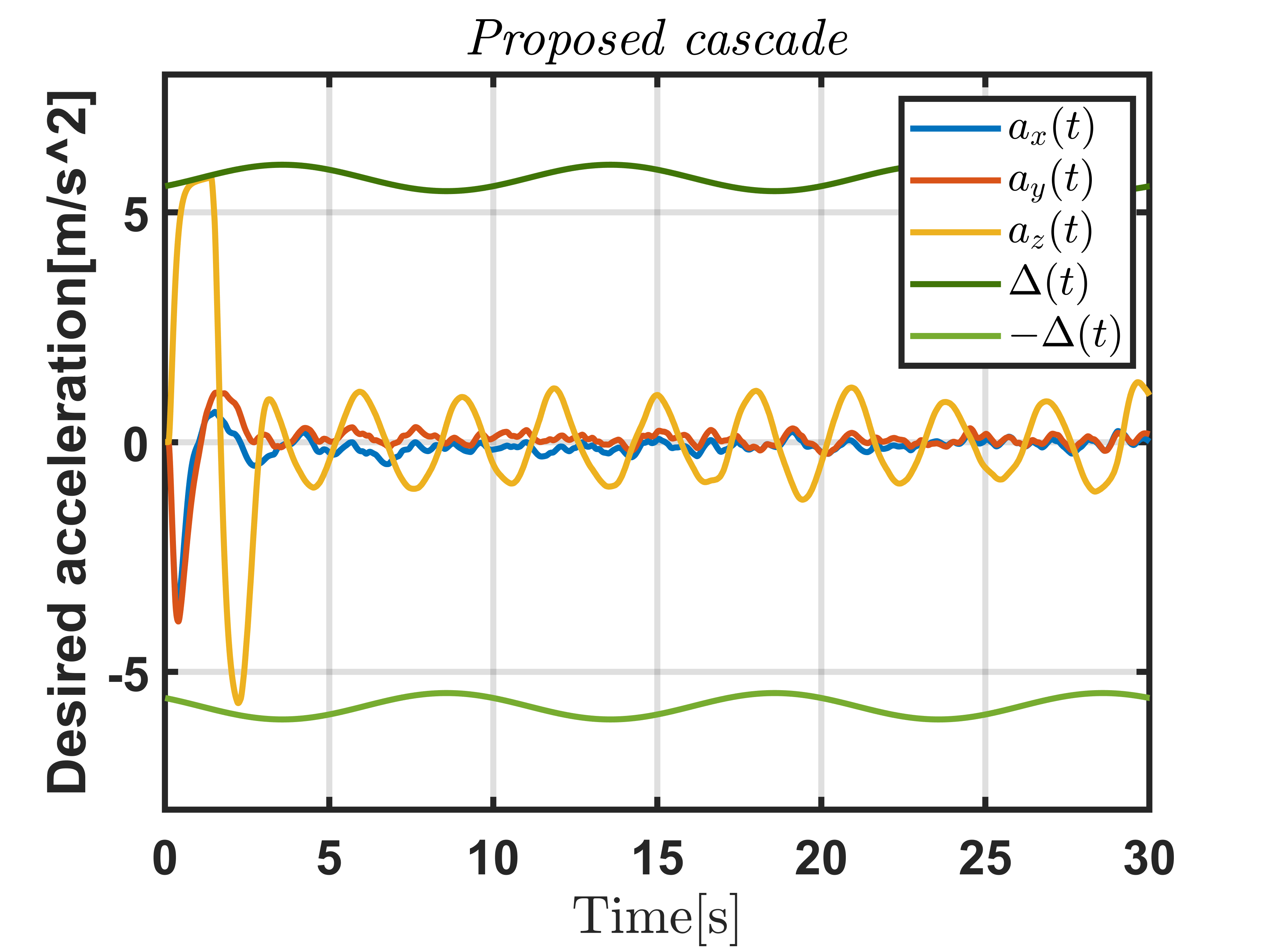}
\caption{Desired accelerations in the $x$, $y$, and $z$ directions for the \textit{Proposed Cascade} in the Avular high-fidelity simulation environment show that the MPC controller keeps $a_{d,i}$ within $\pm \Delta(t)$, satisfying constraint (\ref{TV cons decoupled}).}
\label{ad_sim 2}
\end{figure}

The Root Mean Square Error (RMSE) for the \textit{Proposed cascade} and \textit{PID} are $\unit[[0.17, 0.086, 1.41]]{m}$ and $\unit[[0.20, 0.12, 1.47]]{m}$, respectively, demonstrating the effectiveness of the \textit{Proposed cascade} in tracking fast trajectories with small error. Note that the \textit{proposed cascade} not only outperforms \textit{PID} with a lower RMSE but also offers formal stability guarantees, whereas \textit{PID} does not provide such guarantees.
Fig. \ref{ad_sim 2} shows the desired accelerations in the $x$, $y$, and $z$ directions for \textit{Proposed cascade}, highlighting that the desired acceleration constraints in (\ref{TV cons decoupled}) are met.
%\vspace{-0.2cm}
%\section{Experimental Validation} \label{experiment section}
%\input{Sections/Section11}
\vspace{-0.1cm}
\section{Conclusion} \label{Conclusion section}
\vspace{-0.25cm}
This paper has provided a cascade inner-outer loop control structure for trajectory tracking of a quadcopter, including a formal closed-loop tracking guarantee. The proposed approach has been based on a new MPC-based controller design for the outer loop, accounting for the quadcopter's limited thrust capabilities and leading to high performance through a significantly less conservative design compared to earlier MPC proposals with stability guarantees.
The combination of a UaGAS nonlinear controller in the inner loop and a UGAS MPC-based controller in the outer loop has enabled us to demonstrate UaGAS for the trajectory tracking errors of the entire cascade system.

To validate the proposed control schemes, we have conducted a numerical case study to compare the performance of the proposed strategy with the controller from \cite{Alexpaper}. This comparison has highlighted how the proposed method improves the performance of the cascade control strategy, particularly by reducing conservatism.
We have also validated the proposed cascade control strategy by comparing it with a reference PID-based cascade control system in the Avular high-fidelity  simulation environment within MATLAB, using Avular Vertex One parameters. The results have shown that our proposed strategy outperforms PID-based cascade control system in navigating fast trajectories.
%Finally, a real-world lab experiments have been conducted to validate the performance of the MPC-based controller in outer loop. The results have demonstrated that the proposed MPC strategy delivers rapid convergence with small errors. (MPC with longer horizon)

Future work will include relaxing the constraints on $a_d$ by using the first original constraint from (\ref{main ad constraint}) and reducing the conservatism introduced by using the alternative constraint in (\ref{TV cons coupled}). Particularly, the real-time feasibility of a novel MPC design with a coupled 12th-order model and coupled constraints, running on the quadcopter's on-board hardware platform, should be considered. Another avenue for future research involves conducting real-life experiments to validate the entire cascade scheme.
%that use the nonlinear attitude controller in the inner loop, combined with the MPC-based controller in the outer loop, to validate the entire cascade scheme.
\vspace{-0.2cm}

%\vspace{-0.5cm}
%\begin{ack}                               % Place acknowledgements
%Partially supported by the Roman Senate.  % here.
%\end{ack}

\appendix
%\vspace{-0.2cm}
\section{Appendix A. Proof of Theorem \ref{TV theorem}} \label{TV theorem proof}
\vspace{-0.25cm}
If $ u_{{min}_{i|k}}\leq u_{i|k} \leq u_{{max}_{i|k}}$, then 
\begin{subequations}
\begin{align}
        \Tilde{\Delta}^-_{i|k}\leq &~ u_{i|k} \leq \Tilde{\Delta}^+_{i|k},
        \label{first input const}\\
        \bar{\Delta}^-_{i|k}\leq &~ u_{i|k} \leq \bar{\Delta}^+_{i|k},
        \label{second input const}\\
        -\Delta_{i|k}\leq &~ u_{i|k} \leq \Delta_{i|k}.
        \label{third input const}
   \end{align}
\end{subequations}
By meeting (\ref{first input const}) and (\ref{second input const}), in line with Lemma \ref{ad TV lemma} and Lemma \ref{eta TV lemma}, we can ensure that $|a_{d_{i+1|k}}|\leq \Delta_{i+1|k}$ and $|\eta_{i+1|k}| \leq \Delta_{i+1|k}$ for \( i \in \{0, 1, \dots, N-1\} \).
So, by initializing the controller with $|a_{d_{0|0}}| \leq \Delta_{0|0}$ and $|\eta_{0|0}| \leq \Delta_{0|0}$, we have $|a_{d_{i|k}}| \leq \Delta_{i|k}$ and $|\eta_{i|k}| \leq \Delta_{i|k}$ for all $i \in \{0,1,...,N\}$.

If the lower bounds of these three input constraints (\ref{first input const}-\ref{third input const}) are negative and their upper bounds are positive, we ensure that the intersection of these three input constraint sets is not empty. Hence, to establish a single TV input constraint with a negative lower bound $u_{{min}_{i|k}}$ and a positive upper bound $u_{{max}_{i|k}}$, $\Tilde{\Delta}^-_{i|k}$ and $\Bar{\Delta}^-_{i|k}$
should be negative, while 
$\Tilde{\Delta}^+_{i|k}$ and $\Bar{\Delta}^+_{i|k}$ should be positive (knowing that $\Delta_{i|k}$ is always positive). Considering $|a_{d_{i|k}}| \leq \Delta_{i|k}$ and $|\eta_{i|k}| \leq \Delta_{i|k}$, it follows that
\begin{align}
        |\alpha a_{d_{i|k}}+\beta \eta_{i|k}|\leq(\alpha+\beta)\Delta_{i|k},~
        |\alpha \eta_{i|k}| \leq \alpha \Delta_{i|k}.\label{temp2}
\end{align}
If $h$ and $\gamma$ are chosen such that $\Delta_{i+1|k} > (\alpha+\beta) \Delta_{i|k}$ for all $k \in \mathbb{N}_0$, then we can infer from (\ref{temp2}) that $|\alpha a_{d_{i|k}}+\beta \eta_{i|k}| < \Delta_{i+1|k}$ and $|\alpha \eta_{i|k}|< \Delta_{i+1|k}$.
Using \eqref{deltatilde +}, \eqref{deltatilde -}, \eqref{deltabar +} and \eqref{deltabar -}, it is evident that $\Tilde{\Delta}^-_{i|k}$ and $\Bar{\Delta}^-_{i|k}$ are negative, and 
$\Tilde{\Delta}^+_{i|k}$ and $\Bar{\Delta}^+_{i|k}$ are positive.
Therefore, the input set in (\ref{input constraint}) is non-empty.
\vspace{-0.3cm}
\section{Appendix B. Proof of Lemma \ref{smallest input set lemma}} \label{lemma 5 proof}
\vspace{-0.15cm}
 To prove that $\mathscr{U} \subset \mathscr{U}_{i|k}$, it suffices to show that $-\Delta^*\geq u_{{min}_{i|k}}$ and $\Delta^*\leq u_{{max}_{i|k}}$ at all sample times $k$.
According to (\ref{umin}) and (\ref{umax}), we have 
\vspace{-0.1cm}
\begin{multline*}
  \max\limits_{k} (u_{{min}_{i|k}})=\\
       \max\left(\max\limits_{k} -\Delta(k), \max\limits_{k} \Tilde{\Delta}^-(k), \max\limits_{k} \Bar{\Delta}^-(k) \right),  
\end{multline*}
\vspace{-0.8cm}
\begin{multline*}
   \min\limits_{k} (u_{{max}_{i|k}})= \\ \min\left(\min\limits_{k} \Delta(k),  \min\limits_{k} \Tilde{\Delta}^+(k), \min\limits_{k} \Bar{\Delta}^+ (k)\right). 
\end{multline*}
\vspace{-0.1cm}
Based on (\ref{deltatilde +})
\begin{equation*}
 \Tilde{\Delta}^+_{i|k}(k)=(1-\alpha-\beta)^{-1}[~~~\Delta_{i+1|k}-(\alpha a_{d_{i|k}}+\beta \eta_{i|k})].
 \end{equation*}
 Given $\alpha a_{d_{i|k}}+\beta \eta_{i|k}\leq(\alpha+\beta)\Delta_{i|k}$ from (\ref{temp2}), we get
\begin{equation*}
    \Tilde{\Delta}^+_{i|k}(k)\geq(1-\alpha-\beta)^{-1}[~~~\Delta_{i+1|k}-(\alpha+\beta) \Delta_{i|k}].
\end{equation*}
Minimizing both sides of the inequality over $k$ leads to 
\begin{multline*}
    \min_{k} \Tilde{\Delta}^+(k)\geq \\\underbrace{(1-\alpha-\beta)^{-1} \min_{k}[\Delta(k+1)-(\alpha+\beta)\Delta(k)]}_{\Tilde{\Delta}^+_{min}}.
\end{multline*}
Similarly, using \eqref{deltatilde -}, \eqref{deltabar +}, and \eqref{deltabar -}, and considering (\ref{temp2}),  it can be concluded that
\begin{multline*}
    \min_{k} \Bar{\Delta}^+(k)\geq \underbrace{(1-\alpha)^{-1}\min_{k}[\Delta(k+1)-\alpha \Delta(k)]}_{\bar{\Delta}^+_{min}},
\end{multline*}
\vspace{-1.2cm}
\begin{multline*}
    \max_{k} \Tilde{\Delta}^-(k)\leq \\\underbrace{-(1-\alpha-\beta)^{-1}\min_{k}[\Delta(k+1)-(\alpha+\beta)\Delta(k)]}_{-\Tilde{\Delta}^+_{min}} ,
\end{multline*}
\vspace{-0.8cm}
\begin{multline*}
         \max_{k} \Bar{\Delta}^-(k)\leq \underbrace{-(1-\alpha)^{-1}\min_{k}[\Delta(k+1)-\alpha \Delta(k)]}_{-\bar{\Delta}^+_{min}}.
\end{multline*}
So we have
\begin{align*}
    \min\limits_{k} (u_{{max}_{i|k}})&\geq \min (\Tilde{\Delta}^+_{min} ,~ \bar{\Delta}^+_{min},~\Delta_{min})=\Delta^*,\\
    \max\limits_{k} (u_{{min}_{i|k}})&\leq \max\underbrace{(-\Tilde{\Delta}^+_{min} ,~ -\bar{\Delta}^+_{min},~-\Delta_{min})}_{-\Delta^*},
\end{align*}
which implies $\mathscr{U} \subset \mathscr{U}_{i|k}$ at all sample times $k$.

\section{Appendix C. Proof of the boundedness of the solution for (\ref{position error cascade})-(\ref{eta dynamic cascade})} \label{app1}
Using \eqref{CT system coupled}, the dynamics of (\ref{position error cascade})-(\ref{eta dynamic cascade}) can be written as:
\begin{equation*}
    \Dot{x}(t)=Ax(t)+B u_{MPC}(x(kh))+
    \begin{bmatrix}
        0\\
        I_3\\
        0\\
        0
    \end{bmatrix} TR(I-R_e^{\T})~e_3,
\end{equation*}
for all $t \in [kh,kh+h)$ with $x(t) \in \mathbb{R}^{12}$. Inner-loop closed-loop system has $(R_e(t),\omega_e(t))=(I,0)$ as a uniformly locally exponentially stable (ULES) and UaGAS equilibrium \cite{ErijenNLcontroller}.
Given that $x, y$ and $z$ dynamics can be decoupled and T is globally bounded, only boundedness and stability of the following dynamics needs to be studied:
\begin{equation}
    \Dot{x}(t)=A'x(t)+B' u_{MPC}(x(kh))+
    \begin{bmatrix}
        0\\
        1\\
        0\\
        0
    \end{bmatrix} \phi (t),
    \label{CT decoupled cascade}
\end{equation}
for $t \in [kh,kh+h)$ with $x(t) \in \mathbb{R}^{4}$, $A'$ and $B'$ represent the system matrices of continuous-time dynamics (\ref{decoupled outerloop dynamics}) for a single axis. Since $R(I-R_e^{\T})~e_3$ is ULES, there exist positive constants $\lambda>0$ and $c'>0$, where $c'$ depends on the initial condition $(R_e(t_0),\omega_e(t_0))$, such that $|\phi(t)| \leq c' e^{-\lambda t}$.\\
First step is to show boundedness of states at sample times. The discrete-time dynamics of the cascade system derived from \eqref{CT decoupled cascade} can be expressed as follows:
\begin{multline*}
    x(kh+h)= A_d x(kh)+B_d u_{mpc}(x(kh))  \\
    +\int_{0}^{h} e^{A's} \phi(kh+h-s) \,ds \begin{bmatrix}
        0\\
        1\\
        0\\
        0
    \end{bmatrix},
\end{multline*}
with $A_d$ and $B_d$ defined in \eqref{DT system dynamics}. Knowing that $    \left\| \int_{0}^{h} e^{A's} \phi(kh+h-s) \,ds  \right \| \leq c'' e^{-\lambda k h},$
\iffalse
\begin{equation*}
    \left\| \int_{0}^{h} e^{A's} \phi(kh+h-s) \,ds  \right \| \leq c'' e^{-\lambda k h},
\end{equation*}
\fi
the discrete-time dynamics can be expressed as 
\begin{equation}
    x_{k+1}=f(x_k)+g(x_k,z_k) z_k,
     \label{perturbed dynamic}
\end{equation}
where $f(x_k)=A_d x_k+B_d u_{mpc}(x_k)$, $g(x_k,z_k)$ is bounded and $z_k$ is ULES and UaGAS, satisfying  $\|z_k\| \leq c e^{-\lambda k h}$. As demonstrated in previous sections, the Lyapunov function for the outer-loop problem with dynamics $x_{k+1}=A_d x_k+B_d u_{mpc}(x_k)$ is 
\begin{equation*}
    V(x)=x^{\T} M_q x + \lambda \big(x^{\T} M_c x\big)^{\frac{3}{2}},
\end{equation*}
with $M_q$ ans $M_c$ are positive definite matrices defined in \eqref{LyapFunc}. The Lyapunov function $V(x_k)$ satisfies the following properties:
\begin{itemize}
\item[(i)] $V(f(x_k))-V(x_k) \leq 0$, which implies  $\|f(x_k)\| \leq \|x_k\|$,\label{property i}
\item[(ii)] there exists a positive constant $l$ such that $V(x_k) > l \|x_k\|^2$,\label{property ii}
\end{itemize}

To establish the boundedness of the solution $x_k$ of \eqref{perturbed dynamic}, it suffices to demonstrate the boundedness of the Lyapunov function. Leveraging the properties (i) and (ii) of the $V(x)$, the following inequality is derived for the dynamics formulated in \eqref{perturbed dynamic}:
\begin{align*}
    V(x_{k+1})-&V(x_k)=(f_k+g_k z_k)^{\T}M_q(f_k+g_k z_k) \\
    &+\lambda \big((f_k+g_k z_k)^{\T} M_c (f_k+g_k z_k)\big)^{\frac{3}{2}}-V(x_k)\\
    & \leq 0 + c_1 \|x_k\| \|z_k\| + c_2 \|z_k\|^2 + c_3 \|x_k\|^{\frac{3}{2}} \|z_k\|^{\frac{3}{2}}\\
    &+c_4 \|z_k\|^3.
\end{align*}
Using the property (ii) of the $V(x)$ and knowing that $\|z_k\| \leq c e^{-\lambda k h}$, it is deduced
\begin{align*}
   V(x_{k+1})\leq V(x_k)&+c_5 \sqrt{V(x_k)} e^{-\lambda_1 k h}  \\
    &+c_6 V(x_k) e^{-\lambda_2 k h} + c_7 e^{-\lambda_3 k h},
\end{align*}
where $c_1, c_2, \ldots, c_7 > 0$ and $\lambda_1, \lambda_2, \lambda_3 >0 $. Then $V(x_{k+1})$ can be over-approximated as 
\begin{align*}
   V(&x_{k+1}) < \big(1+c_8e^{-\lambda_4 k h}\big) V(x_k)\\
   &+ 2c_8e^{-\lambda_4 k h} \sqrt{1+c_8e^{-\lambda_4 k h}} \sqrt{V(x_k)}+c_8^2e^{-2\lambda_4 k h}\\
   &= \left(\sqrt{1+c_8e^{-\lambda_4 k h}} \sqrt{V(x_k)} + c_8e^{-\lambda_4 k h}\right)^2,
\end{align*}
where $c_8=\max\left(c_5,c_6,\sqrt{c_7}\right)$ and $\lambda_4=\min\left(\lambda_1, \lambda_2, \frac{\lambda_3}{2} \right)$. Therefore, to demonstrate the boundedness of $V(x_k)$, it suffices to prove that the solution to the following difference equation remains bounded:
\begin{equation*}
    F_{k+1}=  \big(a_k \sqrt{F_k}+b_k\big)^2,
\end{equation*}
where $a_k=\sqrt{1+c_8e^{-\lambda_4 k h}}$ and $b_k=c_8e^{-\lambda_4 k h}$. Considering $F_k=G_k^2$ and given that $a_k, b_k>0$, it is described by the linear difference equation $G_{k+1} = a_k G_k + b_k$, which can be solved iteratively as:
\begin{equation*}
    G_k=A_k G_0 +B_k,~~G(0)=G_0,
\end{equation*}
with $A_k=  \prod_{i=0}^{k-1} a_i, ~~~ B_k= \sum_{j=0}^{k-1} \big( b_j \prod_{i=j+1}^{k-1} a_i \big).$
\iffalse
\begin{equation*}
 A(k)=  \prod_{i=0}^{k-1} a_i, ~~~ B(k)= \sum_{j=0}^{k-1} \big( b_j \prod_{i=j+1}^{k-1} a_i \big).
\end{equation*}
\fi
To demonstrate the boundedness of the solution to the difference equation, it is necessary to prove that $A_k$ and $B_k$ remain bounded. Taking the logarithm of $A_k$:
\begin{align*}
    \log(A_k)&=\log\biggr(\prod_{i=0}^{k-1}\sqrt{1+c_8e^{-\lambda_4 i h}}\biggr)\\
    &=\frac{1}{2} \sum_{i=0}^{k-1} \log\big(1+c_8e^{-\lambda_4 i h}\big)\\
    &\leq \frac{1}{2}\sum_{i=0}^{k-1} c_8e^{-\lambda_4 i h}=   \frac{c_8(1-e^{-\lambda_4 k h})}{2(1-e^{-\lambda_4 h})}\\
    &\leq \frac{c_8}{2(1-e^{-\lambda_4 h})},
    \end{align*}
results in $ A_k\leq e^{~\frac{c_8}{2(1-e^{-\lambda_4 h})}},$
\iffalse
\begin{equation*}
    A(k)\leq e^{~\frac{c_8}{2(1-e^{-\lambda_4 h})}},
\end{equation*}
\fi
which shows the boundedness of $A_k$. In addition,
\begin{equation*}
 \begin{aligned}
    B_k&=\sum_{j=0}^{k-1} \bigg( c_8e^{-\lambda_4 j h} \prod_{i=j+1}^{k-1} \sqrt{1+c_8e^{-\lambda_4 i h}} \bigg) \\
    &\leq \prod_{i=0}^{k-1} \sqrt{1+c_8e^{-\lambda_4 i h}}\sum_{j=0}^{k-1}  c_8e^{-\lambda_4 j h}\\
    &=A_k \frac{c_8 \big(1-e^{-\lambda_4 k h}\big)}{1-e^{-\lambda_4 h}} \leq \frac{c_8 }{1-e^{-\lambda_4 h}}A_k.
\end{aligned}
\end{equation*}
Thus $G_k \leq \left(G_0+\frac{c_8 }{1-e^{-\lambda_4 h}}\right) e^{\frac{c_8}{2\left(1-e^{-\lambda_4 h}\right)}},$
\iffalse
\begin{equation*}
    G(k) \leq \left(G_0+\frac{c_8 }{1-e^{-\lambda_4 h}}\right) e^{\frac{c_8}{2\left(1-e^{-\lambda_4 h}\right)}},
\end{equation*}
\fi
which demonstrates the boundedness of $G_k$. Consequently, this implies the boundedness of $F_k$ and $V(x_k)$, proving that the solution $x_k=[\Tilde{p}_k~ \Tilde{v}_k~ a_{d_k}~ \eta_k]^{\T}$ of \eqref{perturbed dynamic} remains bounded at each sample time. Moreover, continuity ensures boundedness in between sample times in this scenario.
\vspace{-0.3cm}

%\vspace{-0.8cm}
%\vspace{-0.2cm}

%Each appendix must have a short title.
%\section{Some Latin vocabulary}         % Sections and subsections are supported  
                                        % in the appe

\bibliographystyle{plain}        % Include this if you use bibtex 
\bibliography{autosam}           % and a bib file to produce the 
                                 % bibliography (preferred). The
                                 % correct style is generated by
                                 % Elsevier at the time of printing.

%\begin{thebibliography}{99}     % Otherwise use the  
                                 % thebibliography environment.
                                 % Insert the full references here.
                                 % See a recent issue of Automatica 
                                 % for the style.
%  \bibitem[Heritage, 1992]{Heritage:92}
%     (1992) {\it The American Heritage. 
%     Dictionary of the American Language.}
%     Houghton Mifflin Company.
%  \bibitem[Able, 1956]{Abl:56}
%     B.~C.~Able (1956). Nucleic acid content of macroscope. 
%     {\it Nature 2}, 7--9. 
%  \bibitem[Able {\em et al.}, 1954]{AbTaRu:54}   
%     B.~C. Able, R.~A. Tagg, and M.~Rush (1954).
%     Enzyme-catalyzed cellular transanimations.
%     In A.~F.~Round, editor, 
%     {\it Advances in Enzymology Vol. 2} (125--247). 
%     New York, Academic Press.
%  \bibitem[R.~Keohane, 1958]{Keo:58}
%     R.~Keohane (1958).
%     {\it Power and Interdependence: 
%     World Politics in Transition.}
%     Boston, Little, Brown \& Co.
%  \bibitem[Powers, 1985]{Pow:85}
%     T.~Powers (1985).
%     Is there a way out?
%     {\it Harpers, June 1985}, 35--47.

%\end{thebibliography}

\end{document}